\documentclass[12pt]{article}
\pdfoutput=1

\usepackage{amsmath}
\usepackage{graphicx}
\usepackage{xcolor}
\usepackage{framed}
\definecolor{shadecolor}{rgb}{0.90,0.90,0.90}
\usepackage{hyperref}
\usepackage{subcaption}
\usepackage[utf8]{inputenc}

\usepackage{setspace}
\usepackage{amsmath, amssymb, amsthm, float, graphicx}
\numberwithin{equation}{section}

\hypersetup{colorlinks=false, linkcolor=blue, citecolor=red}

\def\beq{\begin{eqnarray}}\def\eeq{\end{eqnarray}}
\def\be{\begin{equation}}\def\ee{\end{equation}}
\def\g{\gamma}

\def\m{\mu}

\def\a{\alpha}
\def\e{\epsilon}
\def\k{\kappa}
\def\b{\beta}
\def\d{\delta}

\def\D{\Delta}
\def\G{\Gamma}
\def\l{\lambda}
\def\pd{\partial}

\def\bz{\bar{z}}

\def\la{\langle}
\def\ra{\rangle}

\def\mo{{\mathcal{O}}}

\def\G{\Gamma}

\usepackage{bm}
\usepackage{bm}

\begin{document}
\title{\bf Anomalous dimensions at finite conformal spin from OPE inversion}
\date{}

\author{Carlos Cardona\footnote{carlosgiraldo@nbi.ku.dk} ~and Kallol Sen\footnote{kallol.sen@ipmu.jp}\\~~~~\\
Niels Bohr International Academy and Discovery Center,\\University of Copenhagen, Niels Bohr Institute\\ Blegamsvej 17, DK-2100 Copenhagen Ø, Denmark
\\ ~~~~\\
 Kavli Institute for the Physics and Mathematics of the Universe,\\
 University of Tokyo, Kashiwa, Chiba 277-8583, Japan\\}
 
\maketitle
\vskip 2cm
\abstract{We compute anomalous dimensions of higher spin operators in Conformal Field Theory at arbitrary space-time dimension by using the OPE inversion formula of \cite{Caron-Huot:2017vep}, both from the position space representation as well as from the integral \textit{viz. Mellin} representation of the conformal blocks. The Mellin space is advantageous over the position space not only in allowing to write expressions agnostic to the space-time dimension, but also in that it replaces tedious recursion relations in terms of simple sums which are easy to perform. We evaluate the contributions of scalar and spin exchanges in the $t-$channel exactly, in terms of higher order Hypergeometric functions. These relate to a particular exchange of conformal spin $\b=\D+J$ in the $s-$channel through the \textit{inversion formula}. Our results reproduce the special cases for large spin anomalous dimension and OPE coefficients obtained previously in the literature.}


\tableofcontents

\onehalfspacing
\section{Introduction and results}
Recent years have seen a resurgence of the bootstrap program boosted by the developments of \cite{Rattazzi:2008pe} on bounding operator dimensions by  imposing crossing symmetry on correlation functions. Subsequent  applications of this techniques \cite{ElShowk:2012ht, Kos:2014bka, El-Showk:2014dwa, Simmons-Duffin:2016wlq} lead to tremendous progress that can be followed by looking into the recent updated reviews on the topic \cite{Rychkov:2016iqz,Simmons-Duffin:2016gjk,Poland:2018epd}. 

Despite the crossing equation is the suitable tool to analyze conformal observables numerically, there are some regions in parameter space that still allow to be explored analytically.  In particular, a great deal of progress has been made by looking at the spectrum of large spin operators  \cite{Komargodski:2012ek,Fitzpatrick:2012yx}. Based on this analysis a successful perturbation theory in spin has been developed \cite{Alday:2015eya, Alday:2015ota,Alday:2015ewa,Kaviraj:2015cxa,Kaviraj:2015xsa} which allows not only to compute anomalous dimension of large spin operators by also to understand universal properties of those operators in generic conformal field theories. One of the striking achievements of  this approach is that, even when a perturbation expansion has been made in inverse powers of large spin, the given expansion can resum to get results at finite smaller values of the spin. The reason why this happens is due to the analyticity in spin of the conformal partial wave expansion recently proved in \cite{Caron-Huot:2017vep} (see also \cite{Simmons-Duffin:2017nub}), where also a powerful inversion formula  has been derived which express the OPE coefficient of a given operator exchange in terms of a convolution of the double discontinuities of the four-point correlation functions across the light-cone branch cuts. This inversion formula is our main tool in this paper to compute the anomalous dimension of the large spin double-twist operators at large but still finite values of the spin, or in other words we show that the inversion formula indeed resums the large spin expansion of the anomalous dimension.  We do this by writing the four-point function in a conformal partial wave expansion in both, position space and Mellin space. A first consideration of the inversion formula in Mellin space have been made recently in \cite{Cardona:2018nnk} which we developed and improve further here. A boostrap approach in Mellin space has been developed  and applied in the works \cite{Gopakumar:2016wkt, Dey:2017fab,Dey:2016mcs,Gopakumar:2016cpb, Golden:2017fip}, where unlike here, crossing symmetry is guaranteed by construction and the bootstrap equations corresponds to conditions that eliminate spurious exchanging operators.

Even though there are closed forms for the conformal blocks in two and four dimension \cite{Dolan:2000uw,Dolan:2000ut,Dolan:2011dv}, that's not the case in general dimension, in particular there is not known closed form in any odd dimension. 
One of the advantages of working in Mellin space is that it is possible to write the conformal partial waves expansion in arbitrary dimension \cite{Mack:2009mi} and we exploit this fact here.  
The most important result of the paper is demonstrated in section \ref{pos} and again in section \ref{mel}. We consider a correlator of the form $\la\mo_1(x_1)\mo_2(x_2)\mo_2(x_3)\mo_1(x_4)\ra$  in the $z\rightarrow0$ limit so that in the $s-$channel, we consider the product of OPEs $\mo_1\times\mo_2$. In the $t-$channel, the decomposition is between the OPE of $\mo_1\times\mo_1$ and $\mo_2\times\mo_2$ (resembles the decompostion of identical scalars). Specifically, in the $s-$channel, we can write, in the $z\rightarrow0$ limit, 
\be
\mathcal{G}_{J,\D}(z,\bz)\equiv\la\mo_1(x_1)\mo_2(x_2)\mo_2(x_3)\mo_1(x_4)\ra_{s-channel}=\sum_{\tau,\b}z^\frac{\tau}{2}C_\tau(\b)G_{J,\Delta}(z,\bar{z}) \,,
\ee
where $\tau=\D_1+\D_2+\g_{12}(\b)$, $\b=\D+J$ is the conformal spin, with $\g_{12}(\b)$ the anomalous dimension. This decomposition, is related to the contribution in the $t-$channel through the inversion formula of \cite{Caron-Huot:2017vep} and gives,
\be
z^\frac{\tau}{2}C_\tau(\b)=\sum_{J,\D}f_{11(J,\D)}f_{22(J,\D)}\k_\b\int_z^1 d\bz \frac{(1-\bz)^{2a}}{\bz^2}k_\b(\bz) {\rm dDisc}\ [\mathcal{G}_{J,\D}(z,\bz)]\,,
\ee
we will review this formula in the next section. Expanding both sides in the $z\rightarrow0$ limit, we obtain two sets of relations for the anomalous dimensions and the corrections to the OPE coefficients corresponding to the coefficient of the $\log z$ terms and the regular term. 

The  contribution of the scalar exchange in the $t-$channel related to a particular operator of a particular conformal spin $\b=\D+J$ in the $s-$channel is given by\footnote{Note that the expressions below agree with the large $\b$ limit, but has additional contributions for finite $\b$. We thank David Simmons-Duffin for pointing this out to us. In general, there are additional terms in \eqref{eqsc} and \eqref{eqsp}, which for four dimensions were considered in \cite{Liu:2018jhs} and we comment below.   }, 
\begin{shaded}
\begin{align}\label{eqsc}
\begin{split}
\g_{12}^{0,\D}(\b)=&-2f_{14\mo}f_{23\mo}\frac{(a-\frac{\D+\tau'}{2})_\frac{\D}{2}(-a-\frac{\D+\tau'}{2})_\frac{\D}{2}}{(\frac{\b-\D-\tau'}{2}-1)_\frac{\D}{2}(\frac{\b+\tau'}{2}+1)_\frac{\D}{2}}\frac{\G(\D)}{\G(\frac{\D}{2})^2}\\
&\times {}_4F_3\bigg[\begin{matrix}1-h+\frac{\D}{2},\frac{\D}{2},\frac{\D+\tau'}{2}+1-a,\frac{\D+\tau'}{2}+1+a\\1+\D-h,2-\frac{\b-\D-\tau'}{2},1+\frac{\b+\D+\tau'}{2}\end{matrix};1\bigg]\,.
\end{split}
\end{align}
\end{shaded}
A generalization of the above expression is the contribution of a spin$-J$ operator in the $t-$channel in given in section \ref{spinex} and to quote,
\begin{shaded}
\begin{align}\label{eqsp}
\begin{split}
\g_{12}^{J,\D}(\b)=& -2f_{14\mo}f_{23\mo}\frac{(a-\frac{\D-J+\tau'}{2})_\frac{\D-J}{2}(-a-\frac{\D-J+\tau'}{2})_\frac{\D-J}{2}}{(\frac{\b-\D+J-\tau'}{2}-1)_\frac{\D-J}{2}(\frac{\b+\tau'}{2}+1)_\frac{\D-J}{2}}\frac{\G(\D+J)}{(d-2)_J\G(\frac{\D+J}{2})^4}\frac{\pi\G(\frac{\D-J}{2})}{\sin\pi(\frac{J-\D}{2})}\\
&\times\sum_{m=0}^J \frac{(-1)^{J-m}A_m(J,\D)}{\G(1+m-J)\G(1+m-\frac{\D+J}{2})(1-h+m+\frac{\D-J}{2})_{J-m}} \\
&\times {}_5F_4\bigg[\begin{matrix}1,\frac{\D-J}{2},1-h+m+\frac{\D-J}{2},\frac{\D-J+\tau'}{2}+1+a,\frac{\D-J+\tau'}{2}+1-a\\ 1+m-J,1+\D-h,2-\frac{\b-\D+J-\tau'}{2},1+\frac{\b+\D-J+\tau'}{2}\end{matrix};1\bigg]\,.
\end{split}
\end{align}
\end{shaded}
In the limit $z\rightarrow0$, these are exact expressions in $\beta$ as long as the anomalous dimension is keep it small (see below). The definiton of $A_m(J,\D)$ is given in \eqref{amj}. Notice that \eqref{eqsp} reduces to \eqref{eqsc} for $J=0$. 

The rest of the paper is organized as follows. In section \ref{invf}, we provide a brief review of the inversion formula of \cite{Caron-Huot:2017vep}. In section \ref{pos}, we computed the large spin anomalous dimension from position space.  In section \ref{mel}, the inversion formula is analyzed from the Mellin (integral) representation point of view and the contributions to the large spin anomalous dimension from the scalar and spin exchanges are computed. In section \ref{match} agreement between the two approaches is shown. Section \ref{spec} discusses some special cases and recover previous results in the literature. We also consider a perturbative expansion in $d-\e$ dimensions for identical scalars. Section \ref{regg} discusses vaguely how the regular terms can be obtained from both the position space and the Mellin space. We end with some discussions in section \ref{conc}. The relevant details of the computations are provided in appendices. Appendix \ref{intrep} discusses the general integral representation for the conformal block and appendix \ref{simpmack} discusses the relevant simplifications of the Mack polynomials in the limit $z\rightarrow0$. Appendix \ref{spinrec} discusses the recursion relations for the general spin$-J$ conformal blocks in position space. 

\section{Inversion formula}\label{invf}
We would like to consider the correlator of four conformal primary scalar operators, which by conformal invariance, is only a function of cross ratios,
\be \label{4p}
\langle {\cal O}_4(x_4)\cdots{\cal O}_1(x_1)\rangle = 
  \frac{1}{(x_{12}^2)^{\frac12(\Delta_1+\Delta_2)}(x_{34}^2)^{\frac12(\Delta_3+\Delta_4)}}
  \left(\frac{x_{14}^2}{x_{24}^2}\right)^{a}
  \left(\frac{x_{14}^2}{x_{13}^2}\right)^{b}
   {\cal G}(z,\bz)
\ee
where $a=\frac12(\Delta_2-\Delta_1)$, $b=\frac12(\Delta_{3}-\Delta_4)$,
and $z$, $\bz$ are conformal cross-ratios given by,
\be
 z\bz=\frac{x_{12}^2x_{34}^2}{x_{13}^2x_{24}^2},\qquad (1-z)(1-\bz) = \frac{x_{23}^2x_{14}^2}{x_{13}^2x_{24}^2}\,.
\ee
The correlator above, can be expand in an operator product expansion when two operators get close to each other. Expanding in terms of the small distance between, say 1 and 2,  we have the following {\it s-channel} expansion,
\be \label{OPE}
{\cal G}(z,\bz) = \sum_{J,\Delta} f_{12{\cal O}}f_{43{\cal O}}\,G_{J,\Delta}(z,\bar{z}) 
\ee
where the sum runs over the exchanged primary operator with spin $J$ and dimension $\Delta$.
The function  $G_{J,\Delta}$ are termed conformal blocks and are eigenfunctions of the quadratic and quartic Casimir invariants of the conformal group.

In even spacetime dimensions, the conformal blocks can be expressed in a closed form in terms of products of hypergeometric functions. They are very well known in two and four dimensions and are given respectively by
\begin{flalign}\label{blocks24}
&& G_{J,\Delta}(z,\bz) &= \frac{k_{\Delta-J}(z)k_{\Delta+J}(\bz)+k_{\Delta+J}(z)k_{\Delta-J}(\bz)}{1+\delta_{J,0}}
\,,
\\
&& G_{J,\Delta}(z,\bz) &= \frac{z\bz}{\bz-z}\big[
 k_{\Delta-J-2}(z)k_{\Delta+J}(\bz)-k_{\Delta+J}(z)k_{\Delta-J-2}(\bz)\big]\,. & 
\end{flalign}
where,
\be
k_{\beta}(z) = \bz^{\beta/2}\,{}_2F_1(\beta/2+a,\beta/2+b,\beta,z)\,.
\ee
Our main tool in this work is the Lorentzian OPE inversion formula recently derivated by Simon Caron-Huot \cite{Caron-Huot:2017vep}\footnote{see also \cite{Simmons-Duffin:2017nub}}, which we will review quickly in this section. 
The starting point is the spectral representation of the OPE \eqref{OPE} expansion given by \cite{Costa:2012cb},
\be\label{spectralexpansion}
 {\cal G}(z,\bz) = 1+
 \sum_{J=0}^\infty \int_{d/2-i\infty}^{d/2+i\infty} \frac{d\Delta}{2\pi i} \,c(J,\Delta)\,F_{J,\Delta}(z,\bz)\,. 
\ee
The contour integral pick up the physical poles associated to the exchange of operators in a OPE expansion and are contained in the function $c(J,\Delta)$.
The function $F_{J,\Delta}$ is given in terms of a linear combination of conformal blocks plus its shadow respectively as,
\be\label{F}
 F_{J,\Delta}(z,\bz) = \frac12\left(G_{J,\Delta}(z,\bz) + \frac{K_{J,d-\Delta}}{K_{J,\Delta}} G_{J,d-\Delta}(z,\bz)\right), 
\ee
with coefficients given by,
\be\label{kappa}
K_{J,\Delta} = \frac{\Gamma(\Delta-1)}{\Gamma\big(\Delta-\tfrac{d}{2}\big)}\kappa_{J+\Delta},
\qquad
\kappa_\beta=\frac{\Gamma\big(\tfrac{\beta}{2}-a\big)\Gamma\big(\tfrac{\beta}{2}+a\big)\Gamma\big(\tfrac{\beta}{2}-b\big)\Gamma\big(\tfrac{\beta}{2}+b\big)}
 {2\pi^2\Gamma(\beta-1)\Gamma(\beta)}\,, 
\ee
and they form a set of orthogonal functions, such as the relation \eqref{spectralexpansion} can be automatically inverted in order to solve for the partial wave coefficients, 
\be\label{partialwavecoeff}
 c(J,\Delta) =\mathcal{N}(J,\Delta)
  \int d^2z\, \mu(z,\bz)\, F_{J,\Delta}(z,\bz)\,{\cal G}(z,\bz)\,, 
\ee
with the normalization factor,
\be\label{normalization}
\mathcal{N}(J,\Delta)=\frac{\Gamma\big(J+\frac{d-2}{2}\big)\Gamma\big(J+\frac{d}{2}\big)K_{J,\Delta}}{
2\pi\,\Gamma(J+1)\Gamma(J+d-2)K_{J,d-\Delta}}\,. 
\ee 
and the conformal invariant measure given by
\be\label{measure}
 \mu(z,\bz) = \left|\frac{z-\bz}{z\bz}\right|^{d-2} \frac{\big((1-z)(1-\bz)\big)^{a+b}}{(z\bz)^2}\,. 
\ee
When going from the Euclidean to the Lorentzian region, the four-point function ${\cal G}(z,\bz)$ develops branch cuts singularities along the lightcone distances between the scalar in the correlator. The idea is then to explore the analytic structure of the partial wave coefficients \eqref{partialwavecoeff} by deforming the contour of integration in such way that trapping the branch cuts with the deformed contour extracts the associated discontinuities. In order to do that, it is necessary to write  the spectral function $F_{J,\Delta}(z,\bz)$ in terms of solutions of the conformal Casimir equations such as the function can be split up in two parts:  a part that vanishes with the proper power law at infinity and another that vanishes in the same way around the origin. Remarkably, it turns out that the particular combination with this property is actually a conformal block with the quantum numbers $\D$ and $J$ swapped (and shifted by $ d-1$), namely ${ G}_{\D+1-d,J+d-1}$. Once the proper spectral representation has been found, one can freely deform the integration contour by trapping the branch-cuts and hence extracting the discontinuities of the four-point function across them. Notice that for a given cross ratio branch cut, there are associated two lightcone distances and therefore, by crossing  a given cross ratio  branch cut, we are actually crossing two lightcone branch cuts, and therefore a double discontinuity. Denoting by ${\rm dDisc}$ the operation of taken that given double discontinuity and the $s-$channel OPE coefficients by,
\be
C(J,\D)=C^t(J,\D)+(-1)^J C^u(J,\D)\,,
\ee
the final result from Caron-Huot is,
\be\label{exactC}
C^t(J,\Delta) = \frac{\kappa_{J+\Delta}}{4}\int_{0}^1 dz d\bz\, \mu(z,\bz)\, G_{\Delta+1-d, J+d-1}(z,\bz)\,{\rm dDisc}\big[{\cal G}(z,\bz)\big]\,.
\ee
The $u$-channel contribution $C^u$ is the same but with operators 1 and 2 interchanged.
In practice, the OPE coefficients can be extracted from the $\bz$ integration as a power expansion in small $z$, since at this limit, the effect of the $z-$integration is only to produce the poles associated to the coefficient under consideration, in the following way: at leading order in small $z$ \eqref{exactC} is approximated by,
\be
C^t(J,\D)=\int_0^1 \frac{dz}{2z}z^\frac{\tau}{2}\,C^t(z,\beta)\,,
\ee
where the following ``generating function'' has been defined,
\be\label{generating}
C^t(z,\beta)\equiv\int_z^1{d\bz\,(1-\bz)^{a+b}\over \bz^2}\kappa_{\beta}\,k_{\beta}(\bz)\text{dDisc}[\mathcal{G}(z,\bz)]\,,
\ee
which at small$-z$ will be given by a power expansion, such as schematically,
\be
C^t(J,\D)\bigg |_\text{poles}=F(J,\D)\int_0^1 \frac{dz}{2z}\,z^{\tau-\tau_0\over 2}={F(J,\D)\over \tau-\tau_0}\,.
\ee
We have defined the usual conformal twist and spin respectively $\tau=J-\D$ and $\b=\D+J$.
In the main body of the paper, we will be interested in study the contributions to \eqref{generating}  coming from a single exchange, so by using the $t-$channel block decomposition of the four-point point function $\mathcal{G}(z,\bz)$ we can compute that contribution from,
\beq\label{genfunc}
C^t(z,\b)|_{J,\Delta}&=&f_{14(J,\D)}f_{23(J,\D)}\k_\b\int_z^1d\bz\ \frac{(1-\bz)^{a+b}}{\bz^2}k_\b(\bz)\text{dDisc}\bigg[\frac{(z\bz)^\frac{\D_3+\D_4}{2}G_{J,\D}(1-z,1-\bz)}{[(1-z)(1-\bz)]^\frac{\D_2+\D_3}{2}}\bigg]\,\nonumber\\
\eeq
where $f_{i\,j(J,\D)}$ corresponds to the three-point function between the external scalars $i$ and $j$ and the exchanging operator. 

In the remaining of this paper we are mainly interested on an equal-dimensions scalar four-point function.  In such case several comments are in order:  The operator exchanges are limited to even spins $J$. The $C^u$ and the $C^t$ coefficients are the same and therefore it is enough to consider only $C^t$. Additionally we would like to consider the $z\to0$ limit in which the conformal blocks dependence on $z$ splits into a singular contribution containing a $\log(z)$ factor and a regular power contribution.

\section{Spinning anomalous dimension at finite $\b$ from cross-ratios space}\label{pos}
In this section we would like to use the formula \eqref{genfunc} to compute the contribution to the  anomalous dimension of large spin operators from a  scalar exchange. We are going to do this in coordinate space and in latter sections also in Mellin space. In both cases, we are able to give exact expressions at finite $\beta$.

\subsection{Scalar exchange}
The scalar conformal block can be written as a doble power expansion \cite{Dolan:2000ut},
\be\label{scalar_exact}
g_{0,\D}(1-z,1-\bz)=\sum_{n,\,m=0}^{\infty} {\left({\D\over2}\right)_m^2\left({\D\over2}\right)_{m+n}^2\over m!n!\left(\D+1-h\right)_m(\D)_{2m+n}}(1-z)^m(1-\bz)^n(1-z\bz)^n\,,
\ee
where $h=d/2$ and should be noticed that we are expanding in the $t-$channel. From this representation we can take the $z\to 0$ limit to obtain,
\beq\label{scalar_exact_zzero}
g_{0,\D}(1-\bz)&=&-\frac{\Gamma (\D)}{\Gamma \left(\frac{\D
   }{2}\right)^2} (1-\bz)^{\D/2} \left(\log (z) \,
   _2F_1\left(\frac{\D}{2},\frac{\D}{2};-\frac{d}{2}+\D +1;1-\bz\right)\right.\nonumber\\
   &&+\left.\log (\bz) \,
   _2F_1\left(\frac{\D}{2},\frac{\D}{2};-\frac{d}{2}+\D +1;1-\bz\right)+2
   \left(\psi ^{(0)}\left(\frac{\D }{2}\right)+\gamma \right)\right)\,.
\eeq
 In this section  we will focus only on the terms accompanying the $\log(z)$ term, which we will refer to as ``the log term'', and we will refer to the remaining terms as ``the regular terms'' which we will consider later. 

As we have mentioned in the section above, at small $z$  the generating function \eqref{genfunc} is given by a power expansion in $z$, whose leading term can be written as,
\be
C^t(z,\b)|_{J,\Delta}\sim C_0(\b) z^{{\tau\over2}+{1\over 2}\gamma_{12}(\beta)}\,.
\ee
If the anomalous dimension $\gamma_{12}(\beta)$ is small,  which is the case  we are going to consider in this work, we can approximate it as,
\be\label{small_gen}
 C^t(z,\b)|_{J,\Delta}\sim z^{{\tau\over2}}\, C_0(\b)\left(1+{1\over 2}\gamma_{12}(\beta)\log(z)\right)\,,
\ee
where $ C_0(\b)$ corresponds to the tree-level OPE square coefficient of the double twist operator corresponding to $\tau=(\D_1+\D_2)$ . 
By comparing the $\log(z)$ term at \eqref{small_gen} with \eqref{scalar_exact_zzero} and using \eqref{genfunc},  the correction to the anomalous dimension $\gamma_{12}(\beta)$ from a scalar exchange is,
\begin{align}\label{anomalouslargejscalar}
\begin{split}
\gamma_{12}(\beta)=&-2f_{14\mo}f_{23\mo}\frac{\G(\D)\k_\b}{\G(\frac{\D}{2})^2 C_0(\b)}\int_0^1 \frac{d\bz(1-\bz)^{\D_2-\D_1}}{\bz^2}k_\b(\bz)\\
&\times {}_2F_1\bigg[{\D\over 2}+a,{\D\over 2}+b,\D-h+1,1-\bz\bigg]\times \text{dDisc}[(1-\bz)^{\frac{\D}{2}-\D_2}\bz^\frac{\D_1+\D_2}{2}]\,.
\end{split}
\end{align}
Here we have taken the ${}_2F_1$ function outside the $\text{dDisc}$ because it is analytic in the argument $1-\bz$. Following \cite{Caron-Huot:2017vep}  in order to perform this integral it is useful to define the following object,
\be\label{Itau}\begin{aligned}
I_{\tau'}^{(a,b)}(\beta)
&\equiv \int_0^1 \frac{d\bz}{\bz^2} (1-\bz)^{a+b}\kappa_\beta k_\beta(\bz)\, {\rm dDisc}\!\left[
\left(\frac{1-\bz}{\bz}\right)^{\frac{\tau'}{2}-b} (\bz)^{-b}\right]
\\ &= 2\sin\pi\left(\frac{\tau'}{2}+a\right)\sin\pi\left(\frac{\tau'}{2}-b\right)\k_\b\frac{\G(\b)\G(\frac{\tau'}{2}+1-b)\G(\frac{\tau'}{2}+1+a)}{\G(\frac{\b}{2}+a)\G(\frac{\b}{2}-b)}\frac{\Gamma\big(\tfrac{\beta}{2}-\tfrac{\tau'}{2}-1\big)}{\Gamma\big(\tfrac{\beta}{2}+\tfrac{\tau'}{2}+1\big)}\,,
\end{aligned}\ee
where the sin($\pi x$) factors comes from taking the double discontinuity on the term in brackets.
The square OPE coefficient $C_0(\b)$ corresponds to taking the tree-level double twist  $\tau'=-\tau_{0}=-(\D_1+\D_2)$, i.e,  $C_0(\b)=I_{-(\D_1+\D_2)}^{(0,0)}(\beta)$,  $\tau_0$  meaning the tree-level twist of the double twist operators.
It is also convenient to use the following transformation of the ${}_2F_1$,
\begin{align}
\begin{split}
&{}_2F_1\bigg[{\D\over2}+a,{\D\over2}+b,\D-h+1,1-\bz\bigg]\\
=&\bz^{-{\D\over2}-a}{}_2F_1\bigg[\frac{\D}{2}+a,\frac{\D-2h}{2}+1-b,\D-h+1,-y\bigg]
\end{split}
\end{align}
with $y={1-\bz\over\bz}$.
 By using the power series expansion of the Gauss hypergeometric in \eqref{anomalouslargejscalar} and using \eqref{Itau} to perform the integral term by term, we arrive to,
\begin{align}\label{scalcont}
\begin{split}
&\int_0^1 \frac{d\bz(1-\bz)^{a+b}}{\bz^2}k_\b(\bz){}_2F_1\bigg[\frac{\D}{2}+a,\frac{\D-2h}{2}+1-b,\D-h+1,-y\bigg]\times \text{dDisc}[(1-\bz)^{\frac{\D+\tau'}{2}-b}\bz^{-\frac{\tau'}{2}}]\\
=& 2\sin\pi\bigg(\frac{\D+\tau'}{2}+a\bigg)\sin\pi\bigg(\frac{\D+\tau'}{2}-b\bigg)\frac{\G(\b)\G(\frac{\D+\tau'}{2}+a+1)\G(\frac{\D+\tau'}{2}-b+1)}{\G(\frac{\b}{2}-b)\G(\frac{\b}{2}+a)}\frac{\G(\frac{\b-\D-\tau'}{2}-1)}{\G(\frac{\b+\D+\tau'}{2}+1)}\\
&\times {}_4F_3\bigg[\begin{matrix}\frac{\D}{2}-h+1,\frac{\D}{2},\frac{\D+\tau'}{2}+a+1,\frac{\D+\tau'}{2}-b+1\\ \D-h+1,2-\frac{\b-\D-\tau'}{2},1+\frac{\b+\D+\tau'}{2}\end{matrix};1\bigg]\,,
\end{split}
\end{align}
where as before we are using the definition $\tau'=-(\D_1+\D_2)$. Hence we can write the contribution to the anomalous dimension coming from the scalar block as,
\begin{shaded}
\begin{align}\label{anomdim}
\begin{split}
\g_{12}^{0,\D}(\b)=&-2f_{14\mo}f_{23\mo}\frac{\sin\pi\bigg(\frac{\D+\tau'}{2}+a\bigg)\sin\pi\bigg(\frac{\D+\tau'}{2}-b\bigg)}{\sin\pi\bigg(\frac{\tau'}{2}+a\bigg)\sin\pi\bigg(\frac{\tau'}{2}-b\bigg)}\frac{(\frac{\tau'}{2}-b+1)_\frac{\D}{2}(\frac{\tau'}{2}+a+1)_\frac{\D}{2}}{(\frac{\b-\D-\tau'}{2}-1)_\frac{\D}{2}(\frac{\b+\tau'}{2}+1)_\frac{\D}{2}}\frac{\G(\D)}{\G(\frac{\D}{2})^2}\\
&\times {}_4F_3\bigg[\begin{matrix}\frac{\D}{2}-h+1,\frac{\D}{2},\frac{\D+\tau'}{2}+a+1,\frac{\D+\tau'}{2}-b+1\\ \D-h+1,2-\frac{\b-\D-\tau'}{2},1+\frac{\b+\D+\tau'}{2}\end{matrix};1\bigg]\,.
\end{split}
\end{align}
\end{shaded}
When all the scalars in the correlator are the same, i.e. $\D_i\equiv\D_0,\,\,\,i=1,\cdots,4$, we should supplement it with $a=b=0$ and $\tau'=-2\D_0=-\tau_0$. Putting these in \eqref{anomdim}, we can find that the anomalous dimensions of double twist operators of the form $\mo_{12}=\mo_1\pd\pd\dots\pd \mo_2$ in the $s-$channel are given by,
\beq\label{anom1}
\g_{12}^{0,\D}(\b)&=&-\frac{\sin\pi\bigg(\frac{\D}{2}-\D_0\bigg)^2}{\sin\pi(-\D_0)^2}\frac{2f_{11\mo}f_{22\mo}\,\G(\D)(1-\D_0)_\frac{\D}{2}^2}{(\frac{\b-\D+2\D_0}{2}-1)_\frac{\D}{2}(\frac{\b-2\D_0}{2}+1)_\frac{\D}{2}\G(\frac{\D}{2})^2}\nonumber\\
&&\,{}_4F_3\bigg[\begin{matrix}\frac{\D}{2}-h+1,\frac{\D}{2},\frac{\D-2\D_0}{2}+1,\frac{\D-2\D_0}{2}+1\\ \D-h+1,2-\frac{\b-\D+2\D_0}{2},1+\frac{\b+\D-2\D_0}{2}\end{matrix};1\bigg]\,.
\eeq
An important observation should be made at this point. In order to compute the integral \eqref{anomalouslargejscalar}, we have used the power series expansion of the hypergeometric function, which is reliable for values of $\bz$ near to one but behave poorly around $\bz=0$. Therefore the result \eqref{anomdim} should be thought as an asymptotic large$-\beta$ expansion determined by the region $\bz\sim 1$. In order to have a result valid for finite values of $\beta$, we need to perform the integral \eqref{anomalouslargejscalar} {\it exactly}. We don't know yet how to perform such exact integration in general dimension $d$, but fortunately this can be done in $d=4$ and $d=2$ which we will consider independently in section (3.3) below \footnote{We want to thank David Simmons-Duffin, Junyu Liu, Eric Perlmutter and Vladimir Rosenhaus, for important clarifications and comments on this issue and for provide us with the exact integration in $d=4$, which have been addressed in \cite{Liu:2018jhs} }.

We have seen that the inversion formula resums the power expansion in $\beta$, as a reflection of the analiticity in spin. As  we have just mention, here we still need to consider that $\beta$ is large enough such that the anomalous dimension is small, unlike the four-dimensional case to be considered below. For the operators we are considering, the anomalous dimension for large enough $\beta $ scales as  \cite{Komargodski:2012ek},
\be\label{perturbative_reg}
\g_{12}^{0,\D}(\b)\sim {f_{11\mo}f_{22\mo}\over\beta^{2(d-2)}}\,.
\ee

\subsection{Spin exchange}
 Let us now consider the contributions to the anomalous dimension $\g_{12}(\b)$ coming from a spin exchange. 
Here we are going to restrict  again to the terms accompanying the singular $\log(z)$  at the leading $z\to0$ region,  namely,
\be\label{logblock}
z^\frac{\D_3+\D_4}{2}G_{J,\D}(1-z,1-\bz) ={1\over 2}z^\frac{\D_3+\D_4}{2}\log(z)g_{J,\D}(1-\bz)+{\rm reg}\,.
\ee
 In terms of $g_{J,\D}$, the generating function \eqref{genfunc} for a particular spin exchange can be written as,
\beq\label{smallzgen}
C(z,\b)|_{J,\D}
\equiv{z^\frac{\D_3+\D_4}{2}\over2}\log(z)\hat{c}_{J,\D}(\b)\,,
\eeq
where we have defined,
\be
\hat{c}_{J,\D}(\b)={f_{14(J,\D)}f_{23(J,\D)}\over2}\k_\b\int_0^1d\bz\ \frac{(1-\bz)^{a+b}}{\bz^2}k_\b(\bz)\text{dDisc}\bigg[\frac{\bz^\frac{\D_3+\D_4}{2}g_{J,\D}(1-\bz)}{(1-\bz)^\frac{\D_2+\D_3}{2}}\bigg]
\ee
In order to perform the integral we can expand $g_{\D,J}$ as a  power series in ${1-\bz\over\bz}\equiv y$,
\be\label{hexp}
g_{J,\D}(y)=y^{\D-J\over2}\sum_{k=0}g_k(J,\D){y}^{k}\,.
\ee
Higher $k-$powers  of $y$ in the above expansion correspond to contributions from the descendant family of the given primary exchange. 
The generating coefficient $\hat{c}_{\D,J}(z,\b)$ can be rewritten as (with $a=b=0$ and $\D_i=\D_0$),
\be\label{genexp}
\hat{c}_{J,\D}(\b)={f_{14(J,\D)}f_{23(J.\D)}\over2}\k_\b\sum_{k=0}g_k(J,\D)\int_0^1d\bz\ \frac{1}{\bz^2}k_\b(\bz)\text{dDisc}\left({\over}y^{\D-J+2k-2\D_0\over2}\right)\,,
\ee
by using \eqref{Itau} and dividing by the identity contribution, we obtain
\be\label{anom_exp}
\g^{J,\D}(\b)={f_{11(J,\D)}f_{22(J,\D)}\over2\, I_{-2\D_0}^{(0,0)}}\k_\b\sum_{k=0}c_k(J,\D,\beta)\,,
\ee
with
\be \label{ck}
c_k(J,\D,\beta)=g_k(J,\D)I_{-(\D-J+2k-2\D_0)}^{(0,0)}(\beta)\,.
\ee
Here is worth noticing that the contribution to the discontinuity from \eqref{hexp} will come only from the primary at $k=0$, since $k\neq0$ is an integer and therefore $y^k$ is a single valued function. This will applies to the remaining cases considered later in this work.

On the other hand $g_{J,\D}(y)$ satisfy recurrence relations of the type considered in \cite{Dolan:2000ut, El-Showk:2014dwa}, however in the small$-z$ limit those recursion are subtle due to the fact that the quartic and the quadratic Casimir mix the leading and the first term in the Taylor expansion in $z$ and we would like to consider the leading $\log(z)$ term only as in \eqref{logblock}. The adequate recursion at small$-z$ can be obtained from the following Casimir equation \cite{Caron-Huot:2017vep}, 
\be\begin{aligned}
 c_4 \,g_{J,\Delta}(z)&=
 \left(\frac{z}{1-z}\right)^{d-2} \big(2D_z -\tilde{Y}-c_2 +2-d\big) \left(\frac{z}{1-z}\right)^{2-d}
 \big(2D_z +\tilde{Y}-c_2\big)g_{J,\Delta}(z)\,, \label{quartic_cross_channel}
\end{aligned}\ee
obtained from the quadratic and quartic Casimirs given respectively by,
\be\begin{aligned}
 C_2&= D_z+D_{\bz}+(d-2)\frac{z\bz}{z-\bz}\left[(1-z)\partial_z-(1-\bz)\partial_{\bz}\right],\\
 C_4&= \left(\frac{z\bz}{z-\bz}\right)^{d-2} (D_z-D_{\bz})\left(\frac{z\bz}{z-\bz}\right)^{2-d}(D_z-D_{\bz})\,.
\label{Casimir_ops}\end{aligned}\ee
whose eigenvalues are,
\be\begin{aligned}
 c_2&= \tfrac12\left[J(J+d-2)+\Delta(\Delta-d)\right],\\
 c_4&= J(J+d-2)(\Delta-1)(\Delta-d+1)\,.\label{Casimirs}
\end{aligned}\ee
By plugging the power series expansion \eqref{hexp} into \eqref{quartic_cross_channel}, we get the following recursion relation for the coefficients $g_{k-1}(J,\Delta)$,
\be\label{recc}
p_{k-1}(\D,J)\, g_{k-1}(J,\Delta)+p_{k-2}(J,\Delta)\, g_{k-2}(J,\Delta)+p_k(J,\Delta)\,g_k(J,\Delta)=0\,,
\ee
with
\beq
p_{k-1}(J,\Delta)&=&-2 \left(d-2 (k+{\tau\over2})\right)^2 \left(d (-\Delta +J+4 (k+{\tau\over2}-1))+\Delta ^2+(J-2) J-4 (k+{\tau\over2})^2+4\right)\nonumber\\
p_{k-2}(J,\Delta)&=&4 \left(k+{\tau\over2}-2\right) \left(-d+k+{\tau\over2}+1\right) \left(2(k+{\tau\over2}-1)-d\right) \left(2 (k+{\tau\over2})-d\right)\nonumber\\
p_k(J,\Delta)&=&\left(-\Delta +J+2 (k+{\tau\over2})\right)  \left(2 d-\Delta +J-2 (k+{\tau\over2}+1)\right) \nonumber\\
&&\,\,\times\left(-d+\Delta +J+2 (k+{\tau\over2})\right) \left(d+\Delta +J-2 (k+{\tau\over2}+1)\right)\,,\nonumber
\eeq
where $\tau=\D-J$ is the usual conformal twist for the exchanged operators.
From the above recurrence relation, we can compute all the coefficients expanding the conformal block \eqref{hexp}, however they become unmanageable large very quickly. Let us display 
the first few coefficients for arbitrary $\D,\,J$ and $d$, for example,
{\small
\beq\label{exp_coeff}
&&g_1(\D,J)=\frac{(d-\Delta +J-2) ((d-2) J-\Delta  (d+2 J-4))}{2 (d-2 (\Delta +1)) (d+2 J-4)}\nonumber \\
&&g_2(\D,J)= \frac{(d-\Delta +J-4) (d-\Delta +J-2)}{8 (d-2 (\Delta +1)) (d-2
   (\Delta +2)) (d+2 J-6) (d+2 J-4) (d-\Delta +J-3)}\nonumber\\
  && \qquad\quad \left[{\over}\Delta ^2 (d+2 J-6) \left(d^2+d (5 J-9)+2 (J-9)
   J+20\right)\right.\nonumber\\
  &&   \qquad\quad\left.+\Delta  \left(-4 (d-3) J^3+((54-7 d) d-100) J^2-2 (d-10) (d-4) (d-3) J+2 (d-6) (d-4)
   (d-3)\right)\right.\nonumber\\
  &&   \qquad\quad\left.+\Delta ^3 (-(d+2 J-6)) (d+2 J-4)+(d-4) (d-2) (J-2) J (d+J-3){\over}\right]\nonumber\\
&&   g_3(\D,J)= -\frac{(d-\Delta +J-6) (d-\Delta +J-2) }{48 (d-2 (\Delta +1)) (d-2 (\Delta +3)) (d+2 J-8) (d+2 J-4)}\,\times\nonumber\\
   &&\qquad\quad\left[\frac{(d-\Delta +J-6) (d (\Delta -J+8)-4 (3 \Delta +8)+2 (\Delta +5) J)}{(d-2 (\Delta +2)) (d+2 J-6)
   (d-\Delta +J-3)}\right.\nonumber\\
   &&\qquad\quad\left.\left({\over}\Delta ^2 (d+2 J-6) \left(d^2+d (5 J-9)+2 (J-9) J+20\right)\right.\right.\nonumber\\
   &&  \qquad\quad\left.\left.+\Delta  \left(-4 (d-3) J^3+((54-7 d) d-100) J^2-2 (d-10)
   (d-4) (d-3) J+2 (d-6) (d-4) (d-3)\right)\right.\right.\nonumber\\
   &&  \qquad\quad\left.\left.+\Delta ^3 (-(d+2 J-6)) (d+2 J-4)+(d-4) (d-2) (J-2) J (d+J-3){\over}\right) \right.\nonumber\\
   &&  \qquad\quad\left.+2 (-\Delta +J-2) (2 d-\Delta +J-8) ((d-4) \Delta +J (-d+2 \Delta +2)){\over}\right]\,.
\eeq}
By using $g_k(\D,J)$ we can then compute all the coefficients in the expansion \eqref{anom_exp}. For example, at leading order in $y$ we have,
{\small
\beq
&&{c_1(\D,J,\beta)\over c_0(\D,J,\beta)}=\frac{\left(J^2 \left(-d+2 \left(J+\tau
   _{\epsilon }\right)+2\right)+2 J \left(d-\left(J+\tau _{\epsilon }\right){}^2-J-\tau _{\epsilon
   }-2\right)+(4-d) \left(J+\tau _{\epsilon }\right){}^2\right)}{2 (d+2 J-4)  \left(d-2 \left(J+\tau _{\epsilon }+1\right)\right) }\nonumber\\
   &&\quad\times\frac{ \Gamma \left(\frac{1}{2} \left(J-\Delta +2 \Delta _0\right)\right){}^2 \Gamma \left(\frac{\beta }{2}+\frac{1}{2}
   \left(-J+\Delta -2 \Delta _0\right)+1\right) \Gamma \left(\frac{\beta }{2}+\frac{1}{2} \left(J-\Delta +2
   \Delta _0-2\right)-1\right)}{ \Gamma \left(\frac{1}{2} \left(J-\Delta +2 \Delta
   _0-2\right)\right){}^2  \Gamma \left(\frac{\beta
   }{2}+\frac{1}{2} \left(-J+\Delta -2 \Delta _0+2\right)+1\right) \Gamma \left(\frac{\beta }{2}+\frac{1}{2}
   \left(J-\Delta +2 \Delta _0\right)-1\right)}\nonumber\\
\eeq}
which should be a good approximation as long as the ratio $\eqref{perturbative_reg}$ is small.
At the large $\beta$ limit it simplifies to,
\be
{c_1(\D,J,\beta)\over c_0(\D,J,\beta)}=-\frac{\left(\Delta -2 \Delta _0-J+2\right){}^2 \left((d+2 J-4) \tau^2+2 J (d+J-3) \tau+2 (d-2) (J-1) J\right)}{2 \beta ^2 (d+2 J-4) \left(d-2 (J+1)-2 \tau\right)}\,,
   \ee
This expression matches previous results  in the literature \cite{Alday:2015ewa, Alday:2015ota, Kaviraj:2015cxa}.
\footnote{Our $g_k$ coefficients are slightly different to the ones from \cite{Alday:2015ewa} $g_1^{here}=A_1^{there}+{(\D-J)\over2}$, because we are expanding the blocks here in \eqref{hexp} is in $y$, whereas \cite{Alday:2015ewa} expand it in $z$.}
Notice that at leading order in large$-\beta$, each coefficient \eqref{ck} (divided by the leading $c_0$) start at $\beta^{-k}$, more precisely,
\be 
{c_k(\D,J,\beta)\over c_0(\D,J,\beta)}\sim g_k(\D,J)\left[\left(1-{J-\D+2\D_0\over2}\right)_k\right]^2\left({2\over\beta}\right)^{2k}\,.
\ee 
therefore at a given order in a $(\beta^{-1})^2$ expansion, we only need a finite number of coefficients. 

The conformal blocks satisfy a recursion relation in spin for fixed $\D$ \cite{Dolan:2000ut, Dolan:2011dv}, hence we can solve a block for spin $J$ from the conformal blocks at spin $J-1$ and $J-2$, or equivalently, we can write a spin $J$ conformal block in terms of linear combinations of scalar blocks \eqref{scalar_exact}, and subsequently the contribution to the anomalous dimension can be similarly be written in terms of linear combinations of the ${}_4F_3$ in \eqref{anomdim}. Even thought this approach will give us closed expression at finite $\beta $, it become very large and tedious even for the lowest values of $J$. We show the simplest $J=1$ block from this procedure in the appendix. We can still however write a closed expression for any spin in four dimension (as well as in two), which we will consider next. 
\subsection{Four dimensions }
In four dimension the recursion discussed above can be resummed into hypergeometric functions, as we already pointed out at equation \eqref{blocks24},
\be
G_{J,\Delta}(1-z,1-\bz) = \frac{(1-z)(1-\bz)}{z-\bz}\big[
 k_{\Delta-J-2}(1-z)k_{\Delta+J}(1-\bz)-k_{\Delta+J}(1-z)k_{\Delta-J-2}(1-\bz)\big]\,. 
\ee
whose leading log-term around $z<<1$ is given by \footnote{In the case of the scalar $J=0$, is straightforward  to see that the linear combination of hypergeometric functions here, simplifies to the one at the log term in \eqref{scalar_exact_zzero}.},
\be\label{logblock}
G_{\Delta,J}(1-z,1-\bz) ={1-\bz\over \bz}\log(z)\,\left({\Gamma(\D-J-2)\over\Gamma\left({\D-J-2\over2}\right)^2}k_{\D+J}(1-\bz)-{\Gamma(\D-J)\over\Gamma\left({\D-J\over2}\right)^2}k_{\D-J-2}(1-\bz)\right) +{\cal O}(z\log z)\,.
\ee
Putting this leading log into the generating function \eqref{genfunc}, we have to perform the following integral,
\beq\k_{{\b}}\int_z^1d\bz\ \frac{1}{\bz^2}k_{{\b}}(\bz)\left(\frac{\bz}{1-\bz}\right)^{\D_0-1},\left({\Gamma(\D-J-2)\over\Gamma\left({\D-J-2\over2}\right)^2}k_{\D+J}(1-\bz)-{\Gamma(\D-J)\over\Gamma\left({\D-J\over2}\right)^2}k_{\D-J-2}(1-\bz)\right)\,.\nonumber\\
\eeq
This integral has been performed very rencently on \cite{Liu:2018jhs}\footnote{Here we are considering an equal scalars correlator and hence we have specialized the formula from \cite{Liu:2018jhs} accordingly.},
\beq
\Omega_{h,h',p}&\equiv&\int_0^1 {d\bz\over \bz^2}\left({\bz\over 1-\bz}\right)^p k_{2 h}(\bz)k_{2 h'}(1-\bz)\nonumber\\
&=&{\Gamma(2h)\Gamma(h'-p+1)^2\Gamma(-h'+h+p-1)\over\Gamma(h)^2\Gamma(h'+h-p+1)}{}_4F_3\bigg[\begin{matrix}h',h',h'-p+1,h'-p+1\\ 2h',h'+h-p+1,h'-h-p+2\end{matrix};1\bigg]\nonumber\\
&+&{\Gamma(2h')\Gamma(h+p-1)^2\Gamma(h'-h-p+1)\over\Gamma(h')^2\Gamma(h'+h-p-1)}{}_4F_3\bigg[\begin{matrix}h,h,h+p+1,h+p+1\\ 2h,h'+h+p-1,-h'+h+p\end{matrix};1\bigg]\,,\nonumber\\
\eeq
which lead us to,
\beq
&&\g^{J,\D}_{12}(\b)=-{k_{{\beta}}(\bz)\over I_{-2\D_0}^{(0,0)}({\beta})}{\sin\pi\bigg(\D_0-\frac{\D-J}{2}\bigg)^2}\nonumber
   \\&&~~
\times\,\left({\Gamma(\D-J-2)\over\Gamma\left({\D-J-2\over2}\right)^2}\Omega_{\b,\b,\D_0-1}-{\Gamma(\b)\over\Gamma\left({\b\over2}\right)^2}\Omega_{\b,\D-J-2,\D_0-1}\right)
\eeq
or more explicitly,
\begin{shaded}
\beq\label{4d}
\g^{J,\D}_{12}(\b)&=&\frac{\Gamma \left(\frac{\b}{2}\right)^2 \Gamma \left(\Delta _0\right){}^2 \Gamma \left(\frac{\b }{2}-\Delta _0+1\right)}{\Gamma (\b) \Gamma \left(\frac{\b
   }{2}+\Delta _0-1\right) \Gamma \left(\frac{\D-J }{2}-\Delta _0+1\right){}^2 \Gamma \left(\Delta _0-\frac{\D-J }{2}\right){}^2}\nonumber
   \\&&~~\times\left({\Gamma(\D-J-2)\over\Gamma\left({\D-J-2\over2}\right)^2}\Omega_{\b,\D+J,\D_0-1}-{\Gamma(\D+J)\over\Gamma\left({\D+J\over2}\right)^2}\Omega_{\b,\D-J-2,\D_0-1}\right)\,.\nonumber\\
\eeq
\end{shaded}
In the Mellin space consideration below, the exact matching with \eqref{4d} or more generally in generic dimensions, requires us to consider the contribution of additional poles in the $s-$integral. In the sections below, we have considered the contribution of the $s=n$ pole which dominates in the large $\b$ limit. When doing the $s-$integral, we should be considering all the contributions (choosing the contour either on {\it rhs} or otherwise) that fall within the contour. The additional poles contributing at finite $\b$ come from $s=(\b-\D+J-\tau)/2-1+n$, which we have neglected in the later calculations.

\section{Spinning anomalous dimension at finite $\b$ from Mellin space}\label{mel}
In the previous sections we have discussed the inversion formula of \cite{Caron-Huot:2017vep} from  position space conformal blocks. This section onwards, we will discuss it by alternative using the integral representation of the conformal blocks \textit{i.e. the Mellin space}. As we will see, working in Mellin space representation have some nice advantages. On one hand, it allow us to write expressions, which are democratic with respect to the space-time dimensions. Even more appealing is that, unlike the cross-ratios conformal blocks in general dimension, we can write a compact representation for the blocks in terms of a contour integral that lately allow us to write them in a power series expansion without the need of solving the cumbersome recursion relations discussed in previous sections.   

Let us start with the definition for the physical block in \eqref{physblck} of appendix \ref{intrep}, 
\begin{align}
\begin{split}
G_{J,\D}(z,\bz)=\frac{\G(\D+J)\G(1+\D-h)}{(d-\D-1)_J\g_{\l_1,a}\g_{\l_1,b}}&\int ds dt\ \frac{\G(\l_2-s)e^{\pi i(\l_2-s)}}{\G(1+s-\bar{\l}_2)}\G(-t)\G(-t-a-b)\G(s+t+a)\\
&\times\G(s+t+b)\mathcal{P}_{J,\D}(s,t,a,b)(z\bz)^s((1-z)(1-\bz))^t\,. 
\end{split}
\end{align}
in complex $(z,\bz)$ coordinates in the $t-$channel. We are considering a subclass of the most general form of correlators: $\la\mo_1\mo_2\mo_2\mo_1\ra$, for which the $t-$channel block has $a=b=0$. See appendix \ref{intrep} for more details. Furthermore, since we are considering $z\rightarrow0$, our starting point for the $t-$channel block is \eqref{tchblock2},  
\begin{align}
\begin{split}
&\lim_{z\rightarrow0}\mathcal{G}^t_{J,\D}(z,\bz)\\
=&z^\frac{\D_1+\D_2}{2}\bigg(\frac{1-\bz}{\bz}\bigg)^{\l_2+\frac{\tau'}{2}-b}\bz^{-b}\frac{\G(\D+J)\G(1+\D-h)}{(d-\D-1)_J\g_{\l_1}^4}\\
&\times\bigg[\log z\sum_{k=0}^\infty \int ds\ (-1)^s\frac{\G(k-s)\G(s-k+\l_2)\G(s+\l_2)}{k!\G(1+s-k+\l_2-\bar{\l}_2)}\bigg(\frac{1-\bz}{\bz}\bigg)^{s}\mathcal{P}_{J,\D}(s-k+\l_2,0,0,0)\\
&+\sum_{k=0}^\infty \int ds\ (-1)^s\frac{\G(k-s)\G(s-k+\l_2)\G(s+\l_2)}{k!\G(1+s-k+\l_2-\bar{\l}_2)}\bigg(\frac{1-\bz}{\bz}\bigg)^{s}\\
&\times\bigg((H(\l_2+s-1)+H(\l_2+s-k-1))\mathcal{P}_{J,\D}(s-k+\l_2,0,0,0)+\mathcal{P}'_{J,\D}(s-k+\l_2,0,0,0)\bigg)\bigg] \,.
\end{split}
\end{align}
where $\tau'=-\D_1-\D_2$ and $b=(\D_2-\D_1)/2$. 

\subsection{Scalar exchange}\label{scalex}

Consider for simplicity the exchange of scalars in the $t-$channel. For this, the Mack polynomial in \eqref{mackp} is $\mathcal{P}_{0,\D}=1$ and the above expression undergoes considerable simplification. Furthermore $\l_2=\D/2$, and, 
\begin{align}
\begin{split}
&\lim_{z\rightarrow0}\mathcal{G}^t_{0,\D}(z,\bz)\\
=&z^\frac{\D_1+\D_2}{2}\bigg(\frac{1-\bz}{\bz}\bigg)^{\frac{\D+\tau'}{2}-b}\bz^{-b}\frac{\G(\D)\G(1+\D-h)}{\G(\frac{\D}{2})^4}\\
&\times\bigg[\log z\sum_{k=0}^\infty \int ds\ (-1)^s\frac{\G(k-s)\G(s-k+\frac{\D}{2})\G(s+\frac{\D}{2})}{k!\G(1+s-k+\D-h)}\bigg(\frac{1-\bz}{\bz}\bigg)^{s}\\
&+\sum_{k=0}^\infty \int ds\ (-1)^s\frac{\G(k-s)\G(s-k+\frac{\D}{2})\G(s+\frac{\D}{2})}{k!\G(1+s-k+\D-h)}\bigg(\frac{1-\bz}{\bz}\bigg)^{s}(H(\D/2+s-1)+H(\D/2+s-k-1))\bigg] \,.
\end{split}
\end{align}
We will come back to the discussion of the regular terms later, but for now, focus on the coefficient of the $\log$ term which contributes to the anomalous dimension of double field operators $\mo_1\pd_{\m_1}\dots\pd_{\m_J}\mo_2$ of dimension $\D=\D_1+\D_2+J+\g_{12}(\b)$ (where $\b=\D+J$ is the conformal spin) in the $s-$channel. The summation over $k$ gives,
\be
\sum_{k=0}^\infty \frac{\G(k-s)\G(s-k+\frac{\D}{2})\G(s+\frac{\D}{2})}{k!\G(1+s-k+\D-h)}=\frac{\G(-s)\G(s+\frac{\D}{2})^2}{\G(1+s+\D-h)}\ {}_2F_1\bigg[\begin{matrix}-s,h-s-\D\\1-s-\frac{\D}{2}\end{matrix};1\bigg]\,.
\ee
Provided we choose to close the contour on the \textit{rhs}, then $\D\geq 2h-2-2s$ is always satisfied due to the unitarity bound. Thus, 
\be
{}_2F_1\bigg[\begin{matrix}-s,h-s-\D\\1-s-\frac{\D}{2}\end{matrix};1\bigg]=\frac{\G(1-s-\frac{\D}{2})\G(1-h+s+\frac{\D}{2})}{\G(1-\frac{\D}{2})\G(1-h+\frac{\D}{2})}\,.
\ee
The coefficient of the $\log$ term becomes, 
\begin{align}
\begin{split}
&\lim_{z\rightarrow0}\mathcal{G}^t_{0,\D}(z,\bz)\bigg|_{\log z}\\
=&z^\frac{\D_1+\D_2}{2}\bigg(\frac{1-\bz}{\bz}\bigg)^{\frac{\D+\tau'}{2}-b}\bz^{-b}\frac{\G(\D)\G(1+\D-h)}{\G(1-\frac{\D}{2})\G(1-h+\frac{\D}{2})\G(\frac{\D}{2})^4}\\
&\times\int ds\ \frac{\G(-s)\G(s+\frac{\D}{2})^2\G(1-s-\frac{\D}{2})\G(1-h+s+\frac{\D}{2})}{\G(1+s+\D-h)}\bigg(-\frac{1-\bz}{\bz}\bigg)^{s}\,.
\end{split}
\end{align}
It is straightforward to see that choosing the poles $s=n$, we can recover the usual log term  scalar block in cross-ratios space, 
\begin{align}
\begin{split}
&\lim_{z\rightarrow0}\mathcal{G}^t_{0,\D}(z,\bz)\bigg|_{\log z}\\
=&-z^\frac{\D_1+\D_2}{2}\bigg(\frac{1-\bz}{\bz}\bigg)^{\frac{\D+\tau'}{2}-b}\bz^{-b}\frac{\G(\D)}{\G(\frac{\D}{2})^2}\ {}_2F_1\bigg[\begin{matrix}\frac{\D}{2},1-h+\frac{\D}{2}\\ \D-h+1\end{matrix};-\frac{1-\bz}{\bz}\bigg]\,,
\end{split}
\end{align}
We will however in this section  use a Mellin space representation of the conformal blocks, which as we will see, allow us to write them in a closed form, unlike the cross-ratio space analysis of sections above, which requires to solve a complicated recursion relation. The idea is to first perform the $\bz$ integral and leave the $s-$integral as the final step to the anomalous dimensions. The resulting coefficient for the $\log$ term, following \eqref{genfunc} and \eqref{Itau} is, 
\begin{align}
\begin{split}
&\int d\bz \frac{(1-\bz)^{2a}}{\bz^2}\k_\b k_\b(\bz)\ {\rm dDisc}\ \bigg[\lim_{z\rightarrow0}\mathcal{G}^t_{0,\D}(z,\bz)\bigg|_{\log z}\bigg]\\
=&2z^\frac{\D_1+\D_2}{2}\frac{\G(\D)\G(1+\D-h)}{\G(1-\frac{\D}{2})\G(1-h+\frac{\D}{2})\G(\frac{\D}{2})^4}\\
&\times\int ds\ (-1)^s\frac{\G(-s)\G(s+\frac{\D}{2})^2\G(1-s-\frac{\D}{2})\G(1-h+s+\frac{\D}{2})}{\G(1+s+\D-h)}I^{(a,a)}_{\D+\tau'+2s}(\b)\,. 
\end{split}
\end{align}
Note that $I^{(a,a)}_{\D+\tau'+2s}(\b)$ has factors $\sin\pi(\frac{\D+\tau'}{2}+s+a)\sin\pi(\frac{\D+\tau'}{2}+s-a)$ coming from the double discontinuity. Since we are choosing the poles of $s=n$ from $\G(-s)$, these factors can be pulled out of the integral in the form of $\sin\pi(\frac{\D+\tau'}{2}+a)\sin\pi(\frac{\D+\tau'}{2}-a)$.  
To obtain the anomalous dimensions, one divides the above expression by the tree-level contribution \textit{i.e.} $I^{(a,a)}_{\tau'}(\b)$ and we obtain,
\begin{align}\label{andimscal}
\begin{split}
\g_{12}^{0,\D}(\b)=&\frac{(a-\frac{\D+\tau'}{2})_\frac{\D}{2}(-a-\frac{\D+\tau'}{2})_\frac{\D}{2}}{(\frac{\b-\D-\tau'}{2}-1)_\frac{\D}{2}(\frac{\b+\tau'}{2}+1)_\frac{\D}{2}}\frac{\G(\D)}{\G(\frac{\D}{2})^2}\\
&\times\int ds\  \G(-s)\frac{(-1)^s(\frac{\D}{2})_s^2(1-h+\frac{\D}{2})_s(\frac{\D+\tau'}{2}+1+a)_s(\frac{\D+\tau'}{2}+1-a)_s}{(1-s-\frac{\D}{2})_s(1+\D-h)_s(\frac{\b+\D+\tau'}{2}+1)_s(\frac{\b-\D-\tau'}{2}-s-1)_s}\,.
\end{split}
\end{align}
Computing the poles of $\G(-s)$ at $s=n$ we can see that, 
\begin{align}
\begin{split}
&\int ds\  \G(-s)\frac{(-1)^s(\frac{\D}{2})_s^2(1-h+\frac{\D}{2})_s(\frac{\D+\tau'}{2}+1+a)_s(\frac{\D+\tau'}{2}+1-a)_s}{(1-s-\frac{\D}{2})_s(1+\D-h)_s(\frac{\b+\D+\tau'}{2}+1)_s(\frac{\b-\D-\tau'}{2}-s-1)_s}\\
=& -{}_4F_3\bigg[\begin{matrix}1-h+\frac{\D}{2},\frac{\D}{2},\frac{\D+\tau'}{2}+1-a,\frac{\D+\tau'}{2}+1+a\\1+\D-h,2-\frac{\b-\D-\tau'}{2},1+\frac{\b+\D+\tau'}{2}\end{matrix};1\bigg]\,,
\end{split}
\end{align}
which matches with that obtained in \eqref{anomdim} for $a=b$. For the sake of completion, we write down the final expression, 
\begin{shaded}
\begin{align}\label{adscal}
\begin{split}
\g_{12}^{0,\D}(\b)=&-\frac{(a-\frac{\D+\tau'}{2})_\frac{\D}{2}(-a-\frac{\D+\tau'}{2})_\frac{\D}{2}}{(\frac{\b-\D-\tau'}{2}-1)_\frac{\D}{2}(\frac{\b+\tau'}{2}+1)_\frac{\D}{2}}\frac{\G(\D)}{\G(\frac{\D}{2})^2}\ {}_4F_3\bigg[\begin{matrix}1-h+\frac{\D}{2},\frac{\D}{2},\frac{\D+\tau'}{2}+1-a,\frac{\D+\tau'}{2}+1+a\\1+\D-h,2-\frac{\b-\D-\tau'}{2},1+\frac{\b+\D+\tau'}{2}\end{matrix};1\bigg]\,.
\end{split}
\end{align}
\end{shaded}

\subsection{Spin exchange}\label{spinex}
A generalization of the scalar exchange is to extend the above formulation to the exchange of spin$-J$ operators in the $t-$channel. We will start with the coefficient of the $\log$ term in \eqref{tchblock2}, 
\begin{align}
\begin{split}
&\lim_{z\rightarrow0}\mathcal{G}^t_{J,\D}(z,\bz)\bigg|_{\log z}\\
=&z^\frac{\D_1+\D_2}{2}\bigg(\frac{1-\bz}{\bz}\bigg)^{\l_2+\frac{\tau'}{2}-b}\bz^{-b}\frac{\G(\D+J)\G(1+\D-h)}{(d-\D-1)_J\g_{\l_1}^4}\\
&\sum_{k=0}^\infty \int ds\ (-1)^s\frac{\G(k-s)\G(s-k+\l_2)\G(s+\l_2)}{k!\G(1+s-k+\l_2-\bar{\l}_2)}\bigg(\frac{1-\bz}{\bz}\bigg)^{s}\mathcal{P}_{J,\D}(s-k+\l_2,0,0,0)\,. 
\end{split}
\end{align}
As explained in appendix \ref{simpmack}, we can use \eqref{plog} to obtain,
\be
\mathcal{P}_{J,\D}(s-k+\l_2,0,0,0)=\frac{(d-\D-1)_J}{(d-2)_J}\sum_{m=0}^J A_m(J,\D)(k-s)_{J-m}\,,
\ee
with $A_m(J,\D)$ given in \eqref{amj}. The coefficient of the $\log$ term then becomes, 
\begin{align}
\begin{split}
&\lim_{z\rightarrow0}\mathcal{G}^t_{J,\D}(z,\bz)\bigg|_{\log z}\\
=&z^\frac{\D_1+\D_2}{2}\bigg(\frac{1-\bz}{\bz}\bigg)^{\frac{\D-J+\tau'}{2}-b}\bz^{-b}\frac{\G(\D+J)\G(1+\D-h)}{(d-2)_J\G(\frac{\D+J}{2})^4}\\
&\sum_{m=0}^J A_m(J,\D)\int ds\ (-1)^s\frac{\G(-s)\G(s+\frac{\D-J}{2})^2}{\G(1+s+\D-h)}\bigg(\frac{1-\bz}{\bz}\bigg)^{s}(-s)_{J-m}\ {}_2F_1\bigg[\begin{matrix}J-m-s,h-\D-s\\ 1-s-\frac{\D-J}{2}\end{matrix};1\bigg]\\
=&z^\frac{\D_1+\D_2}{2}\bigg(\frac{1-\bz}{\bz}\bigg)^{\frac{\D-J+\tau'}{2}-b}\bz^{-b}\frac{\G(\D+J)\G(1+\D-h)}{(d-2)_J\G(\frac{\D+J}{2})^4\G(1-h+\frac{\D+J}{2})}\\
&\sum_{m=0}^J \frac{(-1)^{J-m}}{\G(1+m-\frac{\D+J}{2})} A_m(J,\D)\int ds\ (-1)^s\frac{\G(-s)\G(s+\frac{\D-J}{2})^2}{\G(1+s+\D-h)}\bigg(\frac{1-\bz}{\bz}\bigg)^{s}\\
&\hspace{6cm}\times\frac{\G(1+s)\G(1-s-\frac{\D-J}{2})\G(1-h+m+s+\frac{\D-J}{2})}{\G(1+s+m-J)}\,.
\end{split}
\end{align}
The third line follows from the second line provided we close the contour on the \textit{rhs}, so that the only pole contributions can come from $\G(-s)$ satisfying $1-h+m+s+(\D-J)/2>0$ due to the unitarity bound. Following the discussion in section \ref{scalex}, 
\begin{align}
\begin{split}
\g_{12}^{J,\D}(\b)=&\int_0^1 d\bz \frac{(1-\bz)^{2a}}{\bz^2}\k_\b k_\b(\bz)\ {\rm dDisc}\ \bigg[\lim_{\bz\rightarrow0}\mathcal{G}^t_{J,\D}(z,\bz)\bigg|_{\log z}\bigg]\\
=& 
\frac{\G(\D+J)\G(1+\D-h)}{(d-2)_J\G(\frac{\D+J}{2})^4\G(1-h+\frac{\D+J}{2})}\sum_{m=0}^J \frac{(-1)^{J-m}}{\G(1+m-\frac{\D+J}{2})} A_m(J,\D)\\
&\int ds\ (-1)^s\frac{\G(-s)\G(s+\frac{\D-J}{2})^2\G(1+s)\G(1-s-\frac{\D-J}{2})\G(1-h+m+s+\frac{\D-J}{2})}{\G(1+s+\D-h)\G(1+s+m-J)}\\
&\hspace{10cm}\times I^{(a,a)}_{\D-J+\tau'+2s}(\b)\,,
\end{split}
\end{align}
is the generalization of \eqref{andimscal}, in the case of a spin$-J$ operator exchange in the $t-$channel. We will choose to close the contour on the \textit{rhs} in the complex $s-$plane so that it suffices to consider the poles coming from $\G(-s)$. The poles are at integers $s=n\in\mathbb{I}_{\geq0}$. Thus the $\sin$ factors associated with the ${\rm dDisc}$ can be pulled out of the integral. After some simplifications (and dividing by the tree-level contribution), the above integral can be put in a more convenient form,
\begin{align}
\begin{split}
\g_{12}^{J,\D}(\b)=&
\frac{(a-\frac{\D-J+\tau'}{2})_\frac{\D-J}{2}(-a-\frac{\D-J+\tau'}{2})_\frac{\D-J}{2}}{(\frac{\b-\D+J-\tau'}{2}-1)_\frac{\D-J}{2}(\frac{\b+\tau'}{2}+1)_\frac{\D-J}{2}}\frac{\G(\D+J)}{(d-2)_J\G(\frac{\D+J}{2})^2}\\
&\times\sum_{m=0}^J \frac{(-1)^{J-m}}{\G(1+m-J)} A_m(J,\D)\\
&\int ds\ (-1)^s\frac{\G(-s)(\frac{\D+J}{2})_{s-J}^2(1)_s(1+m-\frac{\D+J}{2})_{J-m-s}(1-h+\frac{\D+J}{2})_{s+m-J}}{(1+\D-h)_s(1+m-J)_s(\frac{\b+\D-J+\tau'}{2}+1)_s(\frac{\b-\D+J-\tau'}{2}-s-1)_s}\\
&\times \bigg(\frac{\D-J+\tau'}{2}+1+a\bigg)_s\bigg(\frac{\D-J+\tau'}{2}+1-a\bigg)_s\,,
\end{split}
\end{align}
where the $s-$integral evaluates to, 
\begin{align}
\begin{split}
&\int ds\ (-1)^s\frac{\G(-s)(\frac{\D+J}{2})_{s-J}^2(1)_s(1+m-\frac{\D+J}{2})_{J-m-s}(1-h+\frac{\D+J}{2})_{s+m-J}}{(1+\D-h)_s(1+m-J)_s(\frac{\b+\D-J+\tau'}{2}+1)_s(\frac{\b-\D+J-\tau'}{2}-s-1)_s}\\
&\times \bigg(\frac{\D-J+\tau'}{2}+1+a\bigg)_s\bigg(\frac{\D-J+\tau'}{2}+1-a\bigg)_s\\
=&\frac{\pi\G(\frac{\D-J}{2})}{\sin\pi(\frac{J-\D}{2})}\frac{1}{\G(\frac{\D+J}{2})^2}\frac{1}{\G(1+m-\frac{\D+J}{2})(1-h+m+\frac{\D-J}{2})_{J-m}}\\
&\times {}_5F_4\bigg[\begin{matrix}1,\frac{\D-J}{2},1-h+m+\frac{\D-J}{2},\frac{\D-J+\tau'}{2}+1+a,\frac{\D-J+\tau'}{2}+1-a\\ 1+m-J,1+\D-h,2-\frac{\b-\D+J-\tau'}{2},1+\frac{\b+\D-J+\tau'}{2}\end{matrix};1\bigg]\,, 
\end{split}
\end{align}
where $a=b=(\D_2-\D_1)/2$. Just for the sake of completion we will write down the final expression as a result of the above simplification,
\begin{shaded}
\begin{align}\label{adspin}
\begin{split}
\g_{12}^{J,\D}(\b)=& 
\frac{(a-\frac{\D-J+\tau'}{2})_\frac{\D-J}{2}(-a-\frac{\D-J+\tau'}{2})_\frac{\D-J}{2}}{(\frac{\b-\D+J-\tau'}{2}-1)_\frac{\D-J}{2}(\frac{\b+\tau'}{2}+1)_\frac{\D-J}{2}}\frac{\G(\D+J)}{(d-2)_J\G(\frac{\D+J}{2})^4}\frac{\pi\G(\frac{\D-J}{2})}{\sin\pi(\frac{J-\D}{2})}\\
&\times\sum_{m=0}^J \frac{(-1)^{J-m}A_m(J,\D)}{\G(1+m-J)\G(1+m-\frac{\D+J}{2})(1-h+m+\frac{\D-J}{2})_{J-m}} \\
&\times {}_5F_4\bigg[\begin{matrix}1,\frac{\D-J}{2},1-h+m+\frac{\D-J}{2},\frac{\D-J+\tau'}{2}+1+a,\frac{\D-J+\tau'}{2}+1-a\\ 1+m-J,1+\D-h,2-\frac{\b-\D+J-\tau'}{2},1+\frac{\b+\D-J+\tau'}{2}\end{matrix};1\bigg]\,.
\end{split}
\end{align}
\end{shaded}
For $J=0$ (and consequently $m=0$), the above formula reduces to \eqref{adscal}. The entire contribution from the $t-$channel can be summarized as,
\be
\g_{12}(\b)=\sum_{\D,J}f_{11(J,\D)}f_{22(J,\D)}\g_{12}^{J,\D}(\b)\,.
\ee

\subsection{Matching cross-ratios conformal blocks}\label{match}
We want to show here that from the previous expressions computed in Mellin space we can recover the coefficients obtained from the conformal blocks in position space.  The coefficients of the $\log$ terms are,
\be
\frac{\G(\D+J)\G(1+\D-h)}{(d-\D-1)_J\G(\frac{\D+J}{2})^4}\sum_{k=0}^\infty\int ds \ \frac{\G(k-s)\G(s-k+\l_2)\G(s+\l_2)}{k!\G(1+s-k+\l_2-\bar{\l}_2)}\bigg(-\frac{1-\bz}{\bz}\bigg)^{s}\mathcal{P}_{J,\D}(s-k+\l_2,0,0,0)\,.
\ee
Using \eqref{plog}, we can write, 
\begin{align}
\begin{split}
&\frac{\G(\D+J)\G(1+\D-h)}{(d-2)_J\G(\frac{\D+J}{2})^4}\sum_{m=0}^J A_m(J,\D)\sum_{k=0}^\infty\int ds \ \frac{\G(k-s)\G(s-k+\l_2)\G(s+\l_2)}{k!\G(1+s-k+\l_2-\bar{\l}_2)}\bigg(-\frac{1-\bz}{\bz}\bigg)^{s} (k-s)_{J-m}\\
=&\frac{\G(\D+J)\G(1+\D-h)}{(d-2)_J\G(\frac{\D+J}{2})^4}\sum_{m=0}^J A_m(J,\D)\int ds \frac{\G(-s)\G(s+\l_2)^2}{\G(1+s+\l_2-\bar{\l}_2)}\bigg(-\frac{1-\bz}{\bz}\bigg)^{s}\\
&\hspace{8cm}\times (-s)_{J-m}\ {}_2F_1\bigg[\begin{matrix}J-m-s,\bar{\l}_2-\l_2-s\\ 1-s-\l_2\end{matrix};1\bigg]\,.
\end{split}
\end{align}
Closing the contour on the \textit{rhs}, one can see that $(\D+2(1+s+m)-d-J)/2>0$ for all $s$ poles and hence,
\be
{}_2F_1\bigg[\begin{matrix}J-m-s,\bar{\l}_2-\l_2-s\\ 1-s-\l_2\end{matrix};1\bigg]=\frac{\G(1-s-\l_2)\G(1+m+s-J-\bar{\l}_2)}{\G(1-J+m-\l_2)\G(1-\bar{\l}_2)}\,.
\ee
The coefficient of the $\log$ term becomes,
\begin{align}
\begin{split}
&\frac{\G(\D+J)\G(1+\D-h)}{(d-2)_J\G(\frac{\D+J}{2})^4}\sum_{m=0}^J A_m(J,\D)\sum_{k=0}^\infty\int ds \ \frac{\G(k-s)\G(s-k+\l_2)\G(s+\l_2)}{k!\G(1+s-k+\l_2-\bar{\l}_2)}\bigg(-\frac{1-\bz}{\bz}\bigg)^{s} (k-s)_{J-m}\\
=&\frac{\G(\D+J)\G(1+\D-h)}{(d-2)_J\G(\frac{\D+J}{2})^4\G(1-h+\frac{\D+J}{2})}\int ds \frac{\G(-s)\G(s+\frac{\D-J}{2})^2\G(1-s-\frac{\D-J}{2})}{\G(1+s+\D-h)}\bigg(-\frac{1-\bz}{\bz}\bigg)^{s}\\
&\hspace{5cm}\times\sum_{m=0}^J \frac{(-1)^{J-m}\G(s+1)\G(1-h+m+s+\frac{\D-J}{2})}{\G(1+m-\frac{\D+J}{2})\G(1+s+m-J)}A_m(J,\D)\\
=&B_0(J,\D)\bigg[1+\sum_{n=0}^\infty g_n(J,\D)\bigg(-\frac{1-\bz}{\bz}\bigg)^n\bigg]\,.
\end{split}
\end{align}
Notice that for a particular $n$ residue, the sum over $m$, can extend from $J$ upto $J-n$ as the terms $m<J-n$ are zero. To fix the normalization, it suffices to evaluate the $n=0$ residue, which gives, 
\be
B_0(J,\D)=-\frac{\G(\D+J)(h-1)_J}{(2h-2)_J\G(\frac{\D+J}{2})^2}\,.
\ee
The coefficients computed previously from position space \eqref{exp_coeff} are then given by $g_i=B_i/B_0$, i.e, with this normalization $g_0=1$ and,
\begin{align}
\begin{split}
g_1(J,\D)=&\frac{(d-\D+J-2)(J(d-2)-\D(d+2J-4))}{2(d+2J-4)(2(\D+1)-d)}\,,\\
g_2(J,\D)=& \frac{(d-\D+J-2)(d-\D+J-4)}{8(d+2J-4)(d+2J-6)(d-2(1+\D))(d-2(2+\D))(d-\D+J-3)}\\
&\times((d-4)(d-2)(J-2)J (d-3+J) + (2 (d-3)(d-4)(d-6) - 2 (d-3)(d-4)(d-10) J \\
&+ ( (54 - 7 d) d-100) J^2 - 4 (d-3) J^3) \D + ( d + 2 J-6) (d^2 + d (5J-9) + 2 (10 + (J-9) J)) \D^2\\
 &- ( d + 2 J-6) ( d + 2 J-4) \D^3)\,,\\
g_3(J,\D)=&\frac{(d-\D+J-2)(d-\D+J-4)(d-\D+J-6)\G(\frac{d}{2}+J-4)\G(\D-\frac{d}{2}+1)}{3072(d-\D+J-3)\G(\frac{d}{2}+J-1)\G(\D-\frac{d}{2}+4)}\\
&\times((-6 + d) (-4 + d) (-2 + d) (-4 + J) (-2 + J) J (-3 + d + 
     J) \\
&- (8 (-8 + d) (-6 + d) (-4 + d) (-3 + d) - 4 (-3 + d) (-16 + 3 d) (46 + (-15 + d) d) J \\
&+ (5024 + 3 d (-1296 + d (342 + (-35 + d) d))) J^2 + 2 (-564 + d (378 + d (-78 + 5 d))) J^3\\
& + 2 (44 + 3 (-8 + d) d) J^4) \D + (-8 + d + 2 J) (-2 (-6 + d) (-4 + d) (-13 + 3 d) \\
&+ (-844 + d (464 + 3 (-25 + d) d)) J + 4 (71 + 3 (-10 + d) d) J^2 + 6 (-4 + d) J^3) \D^2\\
& - (-8 + d + 2 J) (-6 + d + 2 J) (36 + (-13 + d) d - 28 J + 6 d J + 2 J^2) \D^3 \\
&+ (-8 + d + 2 J) (-6 + d + 2 J) (-4 + d + 2 J) \D^4)\,,
\end{split}
\end{align} 
and so on, which of course match the results obtained from the recursion relations, but this time they come from the contour integrals in Mellin space. This is a very non-trivial cross check of our formulas, in particular, even though the compact form for the anomalous dimension \eqref{adspin} looks still complicated, it will be even harder to get to such a formula from the recursion relations in position space, while in Mellin space it boils down simply to the computation of a sum over some residues, which present a clear advantage in comparison with solving algebraic equations.

\section{Special cases}\label{spec}
As some special cases of \eqref{anom1} or equivalently \eqref{adscal}, we will consider the case of identical scalars in the context of the perturbative $\e-$expansion in four dimensions\footnote{See a more detailed account as well higher order corrections to this case in \cite{Alday:2017zzv}}. Furthermore, we also reproduce previous  results obtained  in \cite{Alday:2015ewa, Alday:2015ota, Kaviraj:2015cxa}.  

\subsection{$\e-$expansion for identical scalars}
A special case of \eqref{anom1} is obtained for identical scalars where $\tau=-2\D_\phi$. In that case, we are looking at the anomalous dimensions of operators $\phi\pd\dots\pd\phi$ with $\Delta=2\D_\phi+J+\g_J(\b)$. For the exchange of the scalar $\phi^2$, $\Delta'=2\D_\phi+g\g_{\phi^2}$ and $\D_\phi=h-1+g^2\g_\phi$ in a perturbative expansion in $g$, 
\be
\D+\tau= g\g_{\phi^2}\,.
\ee
We can then write the large spin expansion in the $s-$channel in terms of the low twist scalar exchange in the $t-$channel, given by,
\be
\g_J(\b)=\frac{\G(1-\frac{\e}{2})^2}{\G(-\frac{g\g}{2})^2}\frac{\G(1+g\g-\e)}{\G(\frac{2+g\g-\e}{2})^2}\frac{\G(\frac{\b+\e}{2})\G(\frac{\b-g\g}{2}-1)}{\G(\frac{\b-\e}{2})\G(\frac{\b+g\g}{2}+1)}{}_4F_3\bigg[\begin{matrix}\frac{g\g}{2},1+\frac{g\g}{2},1+\frac{g\g}{2},1+\frac{g\g}{2}\\ 2-\frac{\b-g\g}{2},1+\frac{\b+g\g}{2},1+g\g-\frac{\e}{2}\end{matrix};1\bigg]\,.
\ee
Notice that the expansion begins at $O(g^2)$ because of the $\sin$ factors and the leading order result is,
\be
\lim_{\b\rightarrow\infty}\g_J(\b)= \lim_{\b\rightarrow\infty}\frac{g^2}{4}\g^2 \frac{\G(\frac{\b}{2}-1)}{\G(\frac{\b}{2}+1)}=\frac{g^2}{\b^2}\g^2\,.
\ee
Let's go to the next order. The overall factors, have the expansion,
\begin{align}
\begin{split}
&\frac{\G(1-\frac{\e}{2})^2}{\G(-\frac{g\g}{2})^2}\frac{\G(1+g\g-\e)}{\G(\frac{2+g\g-\e}{2})^2}\frac{\G(\frac{\b+\e}{2})\G(\frac{\b-g\g}{2}-1)}{\G(\frac{\b-\e}{2})\G(\frac{\b+g\g}{2}+1)}\\
&=\frac{g^2\g^2}{\b(\b-2)}\bigg[1+\frac{2g\g}{\b(\b-2)}-(g\g-\e)(1-H(\b/2-1))+O(g^2,\e^2)\bigg]\,.
\end{split}
\end{align}
and the Hypergeometric function can be expanded as,
\begin{align}
\begin{split}
{}_4F_3\bigg[\begin{matrix}\frac{g\g}{2},1+\frac{g\g}{2},1+\frac{g\g}{2},h-1+\frac{g\g}{2}\\ 2-\frac{\b-g\g}{2},1+\frac{\b+g\g}{2},h-1+g\g\end{matrix};1\bigg]=1-\frac{g\g}{2}\bigg(\frac{4}{\b(\b-2)}+\pi\csc\pi\b/2\bigg)\,.
\end{split}
\end{align}
Combining these two, we can write up to $O(g^3,\e^3)$, 
\be
-\frac{g^2\g^2}{\b(\b-2)}\bigg((g\g-\e)(1-H(\b/2-1))+\frac{g\g}{2}\pi\csc\pi\b/2\bigg)\,.
\ee 

\vspace{1cm}

\subsection{Particular dimensions}
In some specific cases the scalar contribution to the anomalous dimension simplifies considerably.  Let us consider some of the cases computed previously in the literature \cite{Alday:2015ewa, Alday:2015ota, Kaviraj:2015cxa}. In order to make the comparison more transparent we set $\tau=-2\D$,  $\D=\D_{\epsilon}$ and $f_{11\mo}f_{22\mo}=f_{0}^2$ in \eqref{anom1}, we can write,
\beq\label{anomalousexactscalar}
\gamma_{12}(\beta,\Delta,\Delta')&=&-2f_{0}^2 \,\frac{\Gamma (\Delta )^2 \Gamma \left(\frac{1}{2} (\beta -2 \Delta +2)\right) \Gamma \left(\Delta _{\epsilon
   }\right) \Gamma \left(\frac{1}{2} \left(\beta +2 \Delta -\Delta _{\epsilon }-2\right)\right)}{\Gamma \left(\frac{1}{2} (\beta +2 \Delta -2)\right) \Gamma \left(\frac{1}{2} \left(2 \Delta
   -\Delta _{\epsilon }\right)\right){}^2 \Gamma \left(\frac{\Delta _{\epsilon }}{2}\right){}^2 \Gamma
   \left(\frac{1}{2} \left(\beta -2 \Delta +\Delta _{\epsilon }+2\right)\right)}\nonumber\\   
   &&{}_4F_3\bigg[\begin{matrix}-\frac{d}{2}+\frac{\Delta _{\epsilon }}{2}+1,\,-\Delta +\frac{\Delta _{\epsilon }}{2}+1,\,-\Delta
   +\frac{\Delta _{\epsilon }}{2}+1,\,\frac{\Delta _{\epsilon }}{2}\\
   -\frac{\beta }{2}-\Delta +\frac{\Delta
   _{\epsilon }}{2}+2,\,\frac{\beta }{2}-\Delta +\frac{\Delta _{\epsilon }}{2}+1,\,-\frac{d}{2}+\Delta _{\epsilon
   }+1\end{matrix};1\bigg]\,,
\eeq

\subsubsection{$d=3,\,\D_{\epsilon}=1$}

The simplest case corresponds to taking $d=3,\,\D_{\epsilon}=1$. 
Plugging it back into \eqref{anomalousexactscalar}, the expression simplifies to
\be 
\gamma_{12}(\beta)=-2f_{0}^2 \,
\frac{\Gamma (\Delta )^2 \Gamma \left(\frac{\beta }{2}-\Delta +1\right) \Gamma \left(\frac{\beta -3}{2}+\Delta \right)}{2 \pi ^2 \Gamma \left(\Delta
   -\frac{1}{2}\right)^2 \Gamma \left(\frac{1}{2} (\beta -2 \Delta +3)\right) \Gamma \left(\frac{\beta }{2}+\Delta -1\right)}
\ee
By further set $\D=1$ and replacing $\beta \to 1-\sqrt{4 j^2+1}$ we got,
\be
\gamma_{12}(\beta)={2\over\sqrt{1 + 4 j^2} \,\pi^3 }\,,
\ee
By Taylor expand around large $j$, we can write the above function as,
\be
\gamma_{12}=-{c_0\over j}\left(1+\sum_{k=1}^{\infty}{c_k\over j^{2k}}\right)\,, 
\ee
where the coefficients of the expansion are given by,
\be
c_k=-\left(\frac{1}{4}\right)^k\frac{ \Gamma \left(k+\frac{1}{2}\right)}{\Gamma (k+1)}\,,
\ee
which is exactly the result quoted in eq. (35)  \cite{Alday:2015ewa}.

\subsubsection{$d=6,\,\D_{\epsilon}=4$}

Now putting the particular values $d=6,\,\D_{\epsilon}=4$ in  \eqref{anomalousexactscalar} we got,
\be
\gamma_{12}=-2f_{0}^2\,\frac{96 \Gamma (\Delta )^2}{ \Gamma (3-\Delta )^2} \frac{1}{(\beta -2 \Delta +2) (\beta -2 \Delta +4) (\beta
   +2 \Delta -6) (\beta +2 \Delta -4)} \,.
\ee
Replacing $\beta \to 1-\sqrt{4 j^2+1}$ in the equation above, we recover the result from 
 eq. (65)  \cite{Alday:2015ewa}.
 \be
\gamma_{12} =-2f_{0}^2\frac{6 \Gamma (\Delta )^2}{j^4 \Gamma (3-\Delta )^2}  \left(1+\frac{j^4}{\left(j^2-(\Delta -3) (\Delta -2)\right)
   \left(j^2-(\Delta -2) (\Delta -1)\right)}\right)
 \ee
 
\section{Regular terms}\label{regg}
The computations considered in sections above only determines the anomalous dimension from the coefficient of the $\log$ terms. As one can see from \eqref{small_gen}, for the OPE coefficients one needs to  analyse  the regular (non-$\log$) terms as well. 

For the scalar block   \eqref{scalar_exact_zzero} the leading regular non-log term is given by,
\beq\label{scalar_block_reg}
&&g_{0,\D'}(1-\bz)=-\frac{\G(\D')}{\G(\frac{\D'}{2})^2}(1-\bz)^{\frac{\D'}{2}}\nonumber\\
&&\times\left({}_2F_1\bigg[\frac{\D'}{2},\frac{\D'}{2},\D'-h+1,1-\bz\bigg]\log(\bz)+2\left(\g+\psi(\D'/2)\right)\right)\,.
\eeq
After plugging it at \eqref{genfunc} and expanding the ${}_2F_1$ function in power series, we need to consider the following complicated integral,
\be\label{Itaulog}\begin{aligned}
\hat{I}_{\tau'}(\beta)
&= \int_0^1 \frac{d\bz}{\bz^2} \kappa_\beta k_\beta(\bz)\, {\rm dDisc}\!\left[
\left(\frac{1-\bz}{\bz}\right)^{\frac{\tau'}{2}} \log(\bz)\right]\,.
\end{aligned}\ee
By further expanding the  $\log$ we can perform the integral,
\beq
\hat{I}_{\tau'}(\beta)&=&\frac{ \Gamma \left(\frac{\beta }{2}\right)^4 \Gamma \left(\frac{1}{2} (\beta -\tau' -2)\right) }{2 \pi ^2  \Gamma (\beta -1) \Gamma (\beta )} \nonumber\\
   &&\sum_{p=1}^{\infty}{(-1)^p\over p}{\Gamma\left(p+\frac{\tau' }{2}+1\right)\over \Gamma \left(p+\frac{\beta }{2}\right)}\, _3F_2\left(\frac{\beta }{2},\frac{\beta }{2},\frac{\beta
   }{2}-\frac{\tau' }{2}-1;p+\frac{\beta }{2},\beta ;1\right)\,.
\eeq
Dividing by the identity, we can write the regular part contribution from the scalar to the coefficient $C_0(\b)$ as, 
\beq
C_{12}^{0,\D'}(\b)=-\frac{\G(\D')}{\G(\frac{\D'}{2})^2\, I_{-2\D_0}^{0,0}}\left({\over}2(\g+\psi(\D'/2))I_{\D'-2\D_0}^{(0,0)}+\sum_{k=0}^{\infty}{\left({\D'\over2}\right)_k^2\over(\D'-h+1)_k }\hat{I}_{\D'-2\D_0+2k}(\beta)\right)
\eeq



We could not find a more compact way to write this expression. In the next section we will consider this contribution from Mellin space.

\subsection{From Mellin space}
We will again start with \eqref{tchblock2} of appendix \ref{intrep} but this time focussing on the non-$\log$ terms. To keep this simple, we will consider the regular terms in the case of scalar exchange. The spin counterpart follows identical logic but with additional complications due to the non-trivial Mack polynomials. The regular terms of \eqref{tchblock2} for scalar exchange in the $t-$channel, are,
\begin{align}\label{regt}
\begin{split}
\lim_{z\rightarrow0}\mathcal{G}^t_{0,\D}(z,\bz)\bigg|_{reg}=&z^\frac{\D_1+\D_2}{2}\bigg(\frac{1-\bz}{\bz}\bigg)^{\frac{\D+\tau'}{2}-b}\bz^{-b}\frac{\G(\D)\G(1+\D-h)}{\G(\frac{\D}{2})^4}\\
&\int ds\ (-1)^s\frac{\G(-s)\G(s+\frac{\D}{2})^2}{\G(1+s+\D-h)}\bigg(\frac{1-\bz}{\bz}\bigg)^{s}\sum_{k=0}^\infty \frac{(-s)_k(1+s-k+\D-h)_k}{k!(s-k+\frac{\D}{2})_k}\\&\times[2\g+\psi(\D/2+s)+\psi(\D/2+s-k)]\,.
\end{split}
\end{align}
The first step is to perform the $k$ sum. This can be done by exploiting an identity,
\be\label{2f1simp}
\sum_{k=0}^\infty \frac{(-s)_k(b-k)_k}{(c-k)_k k!}={}_2F_1\bigg[\begin{matrix}-s,1-b\\ 1-c\end{matrix};1\bigg]=\frac{(b-c)_s}{(1-c)_s}\,, \ \ \mathfrak{Re}(b-c+s)>0\,.
\ee
Closing the contour on the \textit{rhs}, we can see that the last condition is satisfied for $b=1+s+\D-h$ and $c=s+\D/2$, due to the unitarity bound and provided that the exchanged scalar is not a fundamental scalar. Thus, 
\be
\sum_{k=0}^\infty \frac{(-s)_k (1+\D-h+s-k)_k}{(s-k+\frac{\D}{2})_k k!}=\frac{(\frac{\D}{2}-h+1)_s}{(\D-h+1)_s(1-s-\frac{\D}{2})_s}\,.
\ee
Next, the derivative of \eqref{2f1simp}, \textit{wrt} the parameter $c$, gives,
\be
\sum_{k=0}^\infty \frac{(a)_k(b-k)_k}{k!(c-k)_k}(\psi(c-k)-\psi(c))=\frac{(b-c)_s}{(1-c)_s}[\psi(b-c)+\psi(1-c+s)-\psi(1-c)-\psi(b-c+s)]\,.
\ee
which gives,
\be
\sum_{k=0}^\infty \frac{(a)_k(b-k)_k}{k!(c-k)_k}\psi(c-k)=\frac{(b-c)_s}{(1-c)_s}[\psi(c)+\psi(b-c)+\psi(1-c+s)-\psi(1-c)-\psi(b-c+s)]\,.
\ee
For the specified values of $b$ and $c$, we can write,
\begin{align}
\begin{split}
&\sum_{k=0}^\infty \frac{(-s)_k (1+\D-h+s-k)_k}{(s-k+\frac{\D}{2})_k k!}[2\g+\psi(\D/2+s)+\psi(\D/2+s-k)]\\
=&\frac{(\frac{\D}{2}-h+1)_s}{(1-s-\frac{\D}{2})_s}\bigg[2\g+2\psi(s+\D/2)+\psi(\D/2-h+1)+\psi(1-\D/2)\\
&-\psi(1-s-\D/2)-\psi(\D/2-h+1+s)\bigg]\,.
\end{split}
\end{align}
Plugging this back in \eqref{regt}, we find, 
\begin{align}
\begin{split}
&\lim_{z\rightarrow0}\mathcal{G}^t_{0,\D}(z,\bz)\bigg|_{reg}\\
=&z^\frac{\D_1+\D_2}{2}\bigg(\frac{1-\bz}{\bz}\bigg)^{\frac{\D+\tau'}{2}-b}\bz^{-b}\frac{\G(\D)}{\G(\frac{\D}{2})^2}\int ds\ (-1)^s\frac{\G(-s)(\frac{\D}{2})_s^2}{(1+\D-h)_s}\bigg(\frac{1-\bz}{\bz}\bigg)^{s}\frac{(\frac{\D}{2}-h+1)_s}{(1-s-\frac{\D}{2})_s}\\
&\times\bigg[2\g+2\psi(s+\D/2)+\psi(\D/2-h+1)+\psi(1-\D/2)-\psi(1-s-\D/2)-\psi(\D/2-h+1+s)\bigg]\,.
\end{split}
\end{align}
Finally, performing the $\bz$ integral using \eqref{Itau} and dividing by the tree-level contribution, we find, 
\begin{align}
\begin{split}
C^{0,\D}_{12}(\b)=&\frac{(a-\frac{\D+\tau'}{2})_\frac{\D}{2}(-a-\frac{\D+\tau'}{2})_\frac{\D}{2}}{(\frac{\b-\D-\tau'}{2}-1)_\frac{\D}{2}(\frac{\b+\tau}{2}+1)_\frac{\D}{2}}\frac{\G(\D)}{\G(\frac{\D}{2})^2}\\
&\times\int ds\  \G(-s)\frac{(-1)^s(\frac{\D}{2})_s^2(1-h+\frac{\D}{2})_s(\frac{\D+\tau'}{2}+1+a)_s(\frac{\D+\tau'}{2}+1-a)_s}{(1-s-\frac{\D}{2})_s(1+\D-h)_s(\frac{\b+\D+\tau'}{2}+1)_s(\frac{\b-\D-\tau'}{2}-s-1)_s}\\
&\times\bigg[2\g+2\psi(s+\D/2)+\psi(\D/2-h+1)+\psi(1-\D/2)\\
&-\psi(1-s-\D/2)-\psi(\D/2-h+1+s)\bigg]\,,
\end{split}
\end{align}
with $a=(\D_2-\D_1)/2$. We can now consider the $s-$poles from $\G(-s)$ and close the contour on the \textit{rhs}, to obtain,

\begin{shaded}
\begin{align}\label{corrope}
\begin{split}
C^{0,\D}_{12}(\b)=&-\frac{(a-\frac{\D+\tau'}{2})_\frac{\D}{2}(-a-\frac{\D+\tau'}{2})_\frac{\D}{2}}{(\frac{\b-\D-\tau'}{2}-1)_\frac{\D}{2}(\frac{\b+\tau'}{2}+1)_\frac{\D}{2}}\frac{\G(\D)}{\G(\frac{\D}{2})^2}\\
&\times\sum_{n=0}^\infty\frac{(\frac{\D}{2})_n(1-h+\frac{\D}{2})_n(\frac{\D+\tau'}{2}+1+a)_n(\frac{\D+\tau'}{2}+1-a)_n}{n!(1+\D-h)_n(\frac{\b+\D+\tau'}{2}+1)_n(2-\frac{\b-\D-\tau'}{2})_n}\\
&\times\bigg[2\g+2\psi(n+\D/2)+\psi(\D/2-h+1)+\psi(1-\D/2)\\
&-\psi(1-n-\D/2)-\psi(\D/2-h+1+n)\bigg]\,.
\end{split}
\end{align}
\end{shaded}
Although an exact expression is difficult to obtain, one can see that in the large $\b$ limit, the correction can be expanded in the form,
\be
C_{12}^{0,\D}(\b)=-\bigg(\frac{2}{\b}\bigg)^\D\sum_{k=0}^\infty A_k(a,h,\D,\tau')\bigg(\frac{2}{\b}\bigg)^k\,.
\ee
The first few coefficients are of the form,
\begin{align}
\begin{split}
A_0(a,h,\D,\tau')=&2 a_0 H(\D/2-1)\,,\\
A_1(a,h,\D,\tau')=&2a_0\D H(\D/2-1)\,,\\
A_2(a,h,\D,\tau')=&\frac{a_0}{6(\D-h+1)}\bigg[6 (h-1) (2 - 2 a + \D + \tau') (2 + 2 a + \D + \tau') +\D (12 a^2 (2 -2 h + \D)\\ 
&+ (2 + \D) (2 - (\D-4) \D + h (4\D-2)) + 6 (1 + h) \D \tau' + 3 \D \tau'^2) H(\D/2-1)\bigg]\,,
\end{split}
\end{align}
and so on, where,
\be
a_0=\frac{\G(\D)}{\G(\frac{\D}{2})^2}\bigg(a-\frac{\D+\tau'}{2}\bigg)_\frac{\D}{2}\bigg(-a-\frac{\D+\tau'}{2}\bigg)_\frac{\D}{2}\,,
\ee
with $a=(\D_2-\D_1)/2$ and $\tau'=-\D_1-\D_2$.

\subsubsection {Special case: Identical scalars}
We will consider the above non-log term in a special case of identical scalars from the expression in the last subsection. For identical scalars in four dimensions, $a=0$ and $\tau'=-2\D_\phi$. We will consider an $\e-$expansion around the free point, so that $\D=2\D_\phi+g$, and $h=2-\e/2$, and further $\D_\phi=1-\e/2+O(g^2)$. From \eqref{corrope}, we then obtain, for identical scalars, 
\begin{align}
\begin{split}
C_{\phi\phi}^{0,\D}(\b)=&\frac{\G(2h-2+g)}{\G(h-1+\frac{g}{2})^2}\frac{\G(\frac{\b-g}{2}-1)}{\G(\frac{\b+g}{2}+1)}\frac{1}{\G(-\frac{g}{2})^2}\\
&\times \sum_{n=0}^\infty\frac{(h-1+\frac{g}{2})_n (\frac{g}{2})_n (1+\frac{g}{2})_n^2}{n!(h-1+g)_n (\frac{\b+g}{2}+1)_n(2-\frac{\b-g}{2})_n}\bigg[2\g+2\psi\bigg(n+h-1+\frac{g}{2}\bigg)+\psi\bigg(\frac{g}{2}\bigg)\\
&+\psi\bigg(2-h-\frac{g}{2}\bigg)-\psi\bigg(2-h-\frac{g}{2}-n\bigg)-\psi\bigg(n+\frac{g}{2}\bigg)\bigg]\,.
\end{split}
\end{align}
The overall factor can be written as a series expansion in $g$, as follows, 
\begin{align}
\begin{split}
&\frac{\G(2h-2+g)}{\G(h-1+\frac{g}{2})^2}\frac{\G(\frac{\b-g}{2}-1)}{\G(\frac{\b+g}{2}+1)}\frac{1}{\G(-\frac{g}{2})^2}\\
&=\frac{\G(2h-2)}{\G(h-1)^2}\frac{g^2}{\b(\b-2)}\bigg[1-\frac{g}{2}(2H(h-2)-H(2h-3)+H(\b/2-2)+H(\b/2))\bigg]+O(g^3)\,.
\end{split}
\end{align}
Notice that the $n=0$ term of the sum, starts contributing from $O(g,\e)$. The $n=0$ term is simple and,
\be
2H((g-\e)/2)=\frac{\pi^2}{6}(g-\e)-\frac{1}{2}\zeta(3)(g-\e)^2+\frac{\pi^4}{360}(g-\e)^3+O(g^4,\e^4)\,,
\ee
while $n>0$ terms do starting contributing from $O(1)$ and we obtain, 
\begin{align}
\begin{split}
&\sum_{n>0}\frac{(h-1+\frac{g}{2})_n (\frac{g}{2})_n (1+\frac{g}{2})_n^2}{n!(h-1+g)_n (\frac{\b+g}{2}+1)_n(2-\frac{\b-g}{2})_n}\bigg[2\g+2\psi\bigg(n+h-1+\frac{g}{2}\bigg)+\psi\bigg(\frac{g}{2}\bigg)\\
&+\psi\bigg(2-h-\frac{g}{2}\bigg)-\psi\bigg(2-h-\frac{g}{2}-n\bigg)-\psi\bigg(n+\frac{g}{2}\bigg)\bigg]\\
=& \frac{4}{\b(\b-2)}+\pi\csc\pi\b/2+O(g)\,,
\end{split}
\end{align} 
The leading correction to $C_{\phi\phi}^{0,\D}(\b)$ is then,
\be
C_{\phi\phi}^{0,\D}(\b)=\frac{\G(2h-2)}{\G(h-1)^2}\frac{g^2}{\b(\b-2)}\bigg[\frac{4}{\b(\b-2)}+\pi\csc\pi\b/2\bigg]+O(g^3)\,.
\ee

\section{Conclusions and discussion}\label{conc}
In this paper we have computed the anomalous dimension of higher spin operators in conformal field theory by means of the \textit{Inversion Formula} \cite{Caron-Huot:2017vep} both from position space conformal blocks as well as from its Mellin space representation.  In the former case, it is necessary to solve  a recursion relation that computes the coefficients of a power expansion of the conformal blocks in generic dimensions, or equivalent, the coefficients on a expansion in descendant contributions to the anomalous dimension of the spinning operators. In the latter approach,  we are left with an infinite sum which is the same as the left-over one variable integral in the Mellin space. Important distinctions between the two approaches can be observed in the case of a spin$-J$ operator exchange in the $t-$channel. Consider the position space approach first. In this case, we are looking for an expansion of the following sort,
\be
\mathcal{G}^t_{J,\D}(z,\bz)=B_0(J,\D) y^\frac{\D-J}{2}\sum_{k=0}^\infty g_k(J,\D)\ y^k\,,  \ \ y=\frac{1-\bz}{\bz}\,,
\ee
and the coefficients $g_k(J,\D)$ can be obtained through the recursion relations \eqref{recc}. In the case of the (integral) Mellin representation, the recursion relation is replaced by a simple sum over residues. Economically speaking, the sum over terms is much easier to handle than the recursion relation itself. Secondly, the contributions of the scalar/spin exchanges in the $t-$channel can be resummed for any operator in the $s-$channel with finite conformal spin $\b=\D+J$ in terms of general ${}_pF_q$ functions. Thirdly, we have also demonstrated that the formula we obtained in \eqref{adspin} reduces to \eqref{adscal} for $J=0$, and further \eqref{adscal} produces the special cases obtained in \cite{Alday:2015ewa}. Another advantage of the integral representation is taking the $z\rightarrow0$ limit. In terms of the position space representation, taking the $z\rightarrow0$ limit becomes a little cumbersome specially when spin-exchanges are involved. However starting from the (integral) Mellin representation, both the $\log z$ and the regular term can be obtained from the integral representation from the lowest pole in the integral variable. For example we have a following form,
\be
\int dt\ \G(-t)^2 z^t f(t)= f(0)\log z+2\g f(0)+f'(0)\,,
\ee
which is obtained from just the $t=0$ pole of $\G(-t)^2$. By taking the $t=0$ pole, we recover both the $\log$ and the regular term at the same time. The higher orders (away from the $z\rightarrow0$ limit) can be obtained from the $t=n$ poles of $\G(-t)^2$. 

As future perspectives it would be interesting to see how this results relate to previous studies in Mellin space, such as the Mellin bootstrap program \cite{Gopakumar:2016wkt, Dey:2017fab,Gopakumar:2016cpb}. Even more interesting, by considering the Mellin integrand as a scattering process in $AdS$, it would be nice to explore what the results discussed in this paper have to teach us about higher loops corrections to scattering in $AdS$ where some recent considerations have been done in \cite{Cardona:2017tsw, Yuan:2018qva, Yuan:2017vgp, Giombi:2017hpr, Bertan:2018khc} and more generally for the Witten diagrams containing spinning exchanges as the ones considered in \cite{Costa:2014kfa}. Another very attractive follow up is to implement the large spin analysis to correlation functions containing tensorial operators. Such correlations have been study very recently  in Mellin space by \cite{Sleight:2018epi} and it would be very interesting to use those results on a analysis similar to the one performed in this paper. Another perspective might be to explore the inversion formula in the context of Conformal Perturbation Theory recently discussed in \cite{Sen:2017gfr} in general dimensions. One might consider the usefulness of the inversion formula in tying \cite{Sen:2017gfr} together with \cite{Sleight:2018epi} in the context of perturbed $d-$dimensional CFTs, and correlators of general operators. 

\section{Acknowledgments}
We would like to thank  Andrei Parnachev for discussions and pointing out some typos in the first version of this draft. We would also like to thank David Simmons-Duffin for important clarifications and comments on the draft. CC is supported in part by the Danish National Research Foundation (DNRF91), ERC Starting Grant (No 757978) and Villum Fonden. KS would like to thank Niels Bohr Institute, University of Copenhagen, IFT, UAM-CSIC, Madrid and SISSA, Trieste for hospitality during the course of this work. KS is partially supported by WPI initiative, MEXT Japan at IPMU, the University of Tokyo.

\appendix

\section{Integral representation}\label{intrep}

We will start with the integral representation of the conformal blocks following \cite{Dolan:2000ut,Dolan:2011dv}. The integral representation for the four point function $\la\mo_1\mo_2\mo_3\mo_4\ra$ in the OPE decomposition $\mo_1\times\mo_2$ and $\mo_3\times\mo_4$ due to the exchange of an operator $\mo_{J,\D}$, is given by,
\begin{align}\label{intr}
\begin{split}
f_{J,\D}(u,v)=\frac{1}{\g_{\l_1,a}\g_{\bar{\l_1},b}}\int ds dt\ \G(\l_2-s)&\G(\bar{\l}_2-s)\G(-t)\G(-t-a-b)\\
&\times\G(s+t+a)\G(s+t+b)\mathcal{P}_{J,\D}(s,t,a,b)u^s v^t\,,
\end{split}
\end{align}
where we have stripped off the overall kinematical factors. In general,
\be
f_{J,\D}(u,v)=\frac{1}{k_{J,d-\D}}\frac{\g_{\l_1,a}}{\g_{\bar{\l}_1,b}}G_{J,\D}(u,v)+\frac{1}{k_{J,\D}}\frac{\g_{\bar{\l}_1,a}}{\g_{\l_1,a}}G_{J,d-\D}(u,v)\,,
\ee
is a linear combination of the physical block and the shadow respectively from the $s=\l_2+n$ and $s=\bar{\l}_2+n$ poles. To explain, the symbols,
\begin{align}
\l_1=\frac{\D+J}{2}\,, \ \bar{\l}_1=\frac{d-\D+J}{2}\,,\\
\l_2=\frac{\D-J}{2}\,, \ \bar{\l}_2=\frac{d-\D-J}{2}\,.
\end{align}
$a=\frac{\D_{21}}{2}$ and $b=\frac{\D_{34}}{2}$ where $\D_{ij}=\D_i-\D_j$. $(u,v)$ are the conformal cross-ratios, given by,
\be
u=\frac{x_{12}^2x_{34}^2}{x_{13}^2x_{24}^2}=z\bz\,, \ \ v=\frac{x_{14}^2x_{23}^2}{x_{13}^2x_{24}^2}=(1-z)(1-\bz)\,.
\ee
Moreover,
\be
k_{J,\D}=\frac{1}{(\D-1)_J}\frac{\G(d-\D+J)}{\G(\D-h)}\,,\ \ \g_{x,y}=\G(x+y)\G(x-y)\,, \ \text{and}\ h=d/2\,.
\ee
$d-$ is the spacetime dimension. $\mathcal{P}_{J,\D}(s,t,a,b)$ is the Mack polynomial given by,
\begin{align}\label{mackp}
\begin{split}
\mathcal{P}_{J,\D}(s,t,a,b)=&\frac{1}{(d-2)_J}\sum_{m+n+p+q=J}\frac{J!}{m!n!p!q!}(-1)^{p+n}(2\bar{\l}_2+J-1)_{J-q}(2\l_2+J-1)_n(\bar{\l}_1+a-q)_q\\
&\times(\bar{\l}_1+b-q)_q(\l_1+a-m)_m(\l_1+b-m)_m(d-2+J+n-q)_q(h-1)_{J-q}\\
&\times(h-1+n+a+b)_p(\l_2-s)_{p+q}(-t)_n\,.
\end{split}
\end{align}
In order to eliminate the shadow contributions in \eqref{intr} from the start, we will consider a different definition of \eqref{intr}, that produces just the physical blocks. We will write,
\begin{align}
\begin{split}
G_{J,\D}(u,v)=\frac{k_{J,d-\D}}{\g_{\l_1,a}\g_{\l_1,b}}\int ds dt\ F(s)\G(\l_2-s)&\G(\bar{\l}_2-s)\G(-t)\G(-t-a-b)\\
&\times\G(s+t+a)\G(s+t+b)\mathcal{P}_{J,\D}(s,t,a,b)u^s v^t\,,
\end{split}
\end{align}
where $F(s)$ may be thought of as the projection operator\footnote{This is similar to the monodromy operation on the blocks that projects them on to the physical poles.} onto the physical poles. It is not very difficult to see that,
\be
F(s)=\frac{\sin\pi(\bar{\l}_2-s)}{\sin\pi(h-\D)}e^{\pi i (\l_2-s)}\,.
\ee
Combined with this, we can write, 
\begin{align}\label{physblck}
\begin{split}
G_{J,\D}(u,v)=\frac{\G(\D+J)\G(1+\D-h)}{(d-\D-1)_J\g_{\l_1,a}\g_{\l_1,b}}&\int ds dt\ \frac{\G(\l_2-s)e^{\pi i(\l_2-s)}}{\G(1+s-\bar{\l}_2)}\G(-t)\G(-t-a-b)\G(s+t+a)\\
&\times\G(s+t+b)\mathcal{P}_{J,\D}(s,t,a,b)(z\bz)^s((1-z)(1-\bz))^t\,. 
\end{split}
\end{align}
in terms of the complex $z,\bz$ coordinates. In order to simplify matters from the start, we will be dealing with correlators of the form $\la\mo_1\mo_2\mo_2\mo_1\ra$ and investigating the contributions of the $t-$channel exchanges through the inversion formula in \cite{Caron-Huot:2017vep}. For these kind of correlation functions, the $t-$channel contribution essentially reduces to the representation for the identical scalars. The cross-ratios in the $t-$channel, is merely the transformation $(z,\bz)\rightarrow(1-\bz,1-z)$ and with $a=b=0$, we can write,
\begin{align}\label{tchblock}
\begin{split}
G^t_{J,\D}(z,\bz)=\frac{\G(\D+J)\G(1+\D-h)}{(d-\D-1)_J\g_{\l_1}^4}&\int ds dt\ \frac{\G(\l_2-s)e^{\pi i(\l_2-s)}}{\G(1+s-\bar{\l}_2)}\G(-t)^2\G(s+t)^2\\
&\times\mathcal{P}_{J,\D}(s,t,0,0)(z\bz)^t((1-z)(1-\bz))^s\,. 
\end{split}
\end{align}
The above formula will be the starting point of our calculations. We are furthermore interested in the $z\rightarrow0$ limit, where there are simplifications. Before proceeding to the core of the calculations, notice that \eqref{tchblock} is still not in the form most useful for the inversion formula since there are additional factors that we should take into account. The correct quantity in the $t-$channel after taking into account the additional factors is,
\be
\mathcal{G}^t_{J,\D}(z,\bz)=\frac{(z\bz)^\frac{\D_1+\D_2}{2}}{((1-z)(1-\bz))^{\D_2}}G^t_{J,\D}(z,\bz)\,,
\ee
which in the $z\rightarrow0$ limit is,
\begin{align}
\begin{split}
\lim_{z\rightarrow0}\mathcal{G}^t_{J,\D}(z,\bz)=\frac{(z\bz)^\frac{\D_1+\D_2}{2}}{(1-\bz)^{\D_2}}\frac{\G(\D+J)\G(1+\D-h)}{(d-\D-1)_J\g_{\l_1}^4}&\int ds dt\ \frac{\G(\l_2-s)e^{\pi i(\l_2-s)}}{\G(1+s-\bar{\l}_2)}\G(-t)^2\G(s+t)^2\\
&\times\mathcal{P}_{J,\D}(s,t,0,0)z^t \bz^{t+s} \bigg(\frac{1-\bz}{\bz}\bigg)^s\,. 
\end{split}
\end{align}
Now we shift the variable, $s\rightarrow s+\l_2$, so that, 
\begin{align}
\begin{split}
\lim_{z\rightarrow0}z^{-\frac{\D_1+\D_2}{2}}\mathcal{G}^t_{J,\D}(z,\bz)=&\bigg(\frac{1-\bz}{\bz}\bigg)^{\l_2+\frac{\tau}{2}-b}\bz^{-b}\frac{\G(\D+J)\G(1+\D-h)}{(d-\D-1)_J\g_{\l_1}^4}\\
&\times\int ds dt\ (-1)^s\frac{\G(-s)\G(-t)^2\G(s+\l_2+t)^2}{\G(1+s+\l_2-\bar{\l}_2)}\mathcal{P}_{J,\D}(s+\l_2,t,0,0)z^t \bz^{t+s+\l_2} \bigg(\frac{1-\bz}{\bz}\bigg)^s\,. 
\end{split}
\end{align}
where $\tau=-\D_1-\D_2$ and $b=(\D_2-\D_1)/2$ and our contour pick up the poles of $\G(-s)$ at $s=n$. Further, using,
\be
y=\frac{\bz}{1-\bz}\Rightarrow \bz=\frac{1}{1+\frac{1}{y}}\,,\ \text{we write}\ \bz^{s+t+\l_2}=\sum_{k=0}^\infty \frac{(-1)^k}{k!}(s+t+\l_2)_k y^{-k}\,,
\ee
we can write, after further shifting $s\rightarrow s-k$, 
\begin{align}\label{tchblock1}
\begin{split}
\lim_{z\rightarrow0}z^{-\frac{\D_1+\D_2}{2}}\mathcal{G}^t_{J,\D}(z,\bz)=&\bigg(\frac{1-\bz}{\bz}\bigg)^{\l_2+\frac{\tau}{2}-b}\bz^{-b}\frac{\G(\D+J)\G(1+\D-h)}{(d-\D-1)_J\g_{\l_1}^4}\\
&\times\sum_{k=0}^\infty \int ds dt\ (-1)^s\frac{\G(k-s)\G(-t)^2\G(s-k+\l_2+t)\G(s+\l_2+t)}{k!\G(1+s-k+\l_2-\bar{\l}_2)}\\
&\times\mathcal{P}_{J,\D}(s-k+\l_2,t,0,0)z^t \bigg(\frac{1-\bz}{\bz}\bigg)^{s}\,. 
\end{split}
\end{align}
Since we are interested in the $z\rightarrow0$ limit, it only suffices to close the contour on the right and consider the $t=0$ pole. Explicitly,
\begin{align}
\begin{split}
&Res\bigg[\G(-t)^2\G(s-k+\l_2+t)\G(s+\l_2+t)\mathcal{P}_{J,\D}(s-k+\l_2,t,0,0)z^t\bigg]_{t=0}\\
=&\G(s+\l_2)\G(s-k+\l_2)[(\log z+H(\l_2+s-1)+H(\l_2+s-k-1))\mathcal{P}_{J,\D}(s-k+\l_2,0,0,0)\\
&\hspace{11cm}+\mathcal{P}'_{J,\D}(s-k+\l_2,0,0,0)]\,.
\end{split}
\end{align}
where,
\begin{align}
\begin{split}
\mathcal{P}_{J,\D}(s-k+\l_2,0,0,0)=&\frac{1}{(d-2)_J}\sum_{m+p+q=J}\frac{J!}{m!p!q!}(-1)^{p}(2\bar{\l}_2+J-1)_{J-q}(\bar{\l}_1-q)^2_q\\
&\times(\l_1-m)^2_m(d-2+J-q)_q(h-1)_{J-q}(h-1)_p(k-s)_{p+q}\,,\\
\mathcal{P}'_{J,\D}(s-k+\l_2,0,0,0)=&\frac{(1-\d_{n,0})}{(d-2)_J}\sum_{m+n+p+q=J}\frac{J!(n-1)!}{m!n!p!q!}(-1)^{p+n}(2\bar{\l}_2+J-1)_{J-q}(2\l_2+J-1)_n(\bar{\l}_1-q)^2_q\\
&\times(\l_1-m)^2_m(d-2+J+n-q)_q(h-1)_{J-q}(h-1+n)_p(\l_2-s)_{p+q}\,.
\end{split}
\end{align}
The entire contribution from \eqref{tchblock1} can be decomposed into, 
\begin{align}\label{tchblock2}
\begin{split}
&\lim_{z\rightarrow0}\mathcal{G}^t_{J,\D}(z,\bz)\\
=&z^\frac{\D_1+\D_2}{2}\bigg(\frac{1-\bz}{\bz}\bigg)^{\l_2+\frac{\tau}{2}-b}\bz^{-b}\frac{\G(\D+J)\G(1+\D-h)}{(d-\D-1)_J\g_{\l_1}^4}\\
&\times\bigg[\log z\sum_{k=0}^\infty \int ds\ (-1)^s\frac{\G(k-s)\G(s-k+\l_2)\G(s+\l_2)}{k!\G(1+s-k+\l_2-\bar{\l}_2)}\bigg(\frac{1-\bz}{\bz}\bigg)^{s}\mathcal{P}_{J,\D}(s-k+\l_2,0,0,0)\\
&+\sum_{k=0}^\infty \int ds\ (-1)^s\frac{\G(k-s)\G(s-k+\l_2)\G(s+\l_2)}{k!\G(1+s-k+\l_2-\bar{\l}_2)}\bigg(\frac{1-\bz}{\bz}\bigg)^{s}\\
&\times\bigg((H(\l_2+s-1)+H(\l_2+s-k-1))\mathcal{P}_{J,\D}(s-k+\l_2,0,0,0)+\mathcal{P}'_{J,\D}(s-k+\l_2,0,0,0)\bigg)\bigg] \,, 
\end{split}
\end{align}
a $\log$ term, which contributes to the \textit{anomalous dimension} and a finite piece, which contributes to the \textit{correction of the OPE coefficient}. 

\subsection{Simplification of the Mack polynomials}\label{simpmack}
Under certain circumstances, we can simplify the Mack polynomials appearing in \eqref{tchblock2} furthermore. For example, in the coeffiicient of the $\log$ term, one can simply write, 
\begin{align}\label{plog}
\begin{split}
\mathcal{P}_{J,\D}(s-k+\l_2,0,0,0)=&\frac{1}{(d-2)_J}\sum_{m+p+q=J}\frac{J!}{m!p!q!}(-1)^{p}(2\bar{\l}_2+J-1)_{J-q}(\bar{\l}_1-q)^2_q\\
&\times(\l_1-m)^2_m(d-2+J-q)_q(h-1)_{J-q}(h-1)_p(k-s)_{p+q}\,,\\
=&\frac{(d-\D-1)_J}{(d-2)_J} \sum_{m=0}^J A_m(J,\D)(k-s)_{J-m}\,,
\end{split}
\end{align}
where,
\begin{align}\label{amj}
\begin{split}
A_m(J,\D)=&2^{2-2h+\D-J}\sqrt{\pi}\frac{\G(\frac{2h+J-\D}{2})\G(2h-1+m-\D)(h-1)_m}{\G(m+1)\G(\frac{2h-1+J-\D}{2})\G(\frac{2h+2m-J-\D}{2})^2}(1+J-m)_m \bigg(\frac{\D+J}{2}-m\bigg)_m^2\\
&\times (2h-2+m)_{J-m}\ {}_4F_3\bigg[\begin{matrix}m-J,h-1,h+m-1,2h-1+m-\D\\ 2h-2+m,\frac{2h+2m-J-\D}{2},\frac{2h+2m-J-\D}{2}\end{matrix};1\bigg]\,.
\end{split}
\end{align}

\section{Recursion in spin}\label{spinrec}
In general dimension, the conformal blocks satisfies a recursion relation in spin of the form \cite{Dolan:2000ut, Dolan:2011dv},
\beq
&& (\ell + d - 2) \, G_{\Delta,\,(\ell+1)} (a,b;u,v) \nonumber\\
&= & (\ell + {d\over2} -1) \, u^{- {1\over 2}}  \big ( 
G_{\Delta\,(\ell)} (a-{1\over2},b+{1\over2} ;u,v) - 
v G_{\Delta,\,(\ell)} (a+{1\over2},b+{1\over2} ;u,v) \nonumber\\
&&+ G_{\Delta,\,(\ell)} (a+{1\over2},b-{1\over2} ;u,v)
- G_{\Delta,\,(\ell)} (a-{1\over2},b-{1\over2} ;u,v) \big ) - \ell\,  G_{\Delta,\,(\ell-1)} (a,b;u,v) \, . 
\eeq

   In the $t-$channel, one have at $z\to 0$ that coordinates $(u,v)$ go to $u\to (1-\bz),\,v\to 0$, therefore in the limit of interest, the recursion relation can be approximated by
\beq
&& (\ell + d - 2) \, G_{\Delta,\,(\ell+1)} (a,b;(1-\bz),v) \nonumber\\
&= & (\ell + {d\over2} -1) \, u^{- {1\over 2}}  \big ( 
G_{\Delta\,(\ell)} (a-{1\over2},b+{1\over2} ;(1-\bz),v) + G_{\Delta,\,(\ell)} (a+{1\over2},b-{1\over2} ;(1-\bz),v)\nonumber\\
&&
- G_{\Delta,\,(\ell)} (a-{1\over2},b-{1\over2} ;(1-\bz),v) \big ) - \ell\,  G_{\Delta,\,(\ell-1)} (a,b;(1-\bz),v) \, . 
\eeq
   In the small$-z$ limit, the recursion is seeded by the scalar block,
   \be
G_{\Delta,\,(0)} (a,b ;(1-\bz),v)={1\over 2}\log(z)\,(1-\bz)^\frac{\D'}{2}{}_2F_1\bigg[\frac{\D'}{2}+a,\frac{\D'}{2}+b,\D'-h+1,1-\bz\bigg]\,.
\ee
For example $\ell=1$ will be given by
\beq
&& \, g_{\Delta,\,(1)} (a,b;(1-\bz),v)= \nonumber\\
&& {1\over2}\,(1-\bz)^{\frac{\D'}{2}- {1\over 2}} \,  \left( 
{}_2F_1\bigg[\frac{\D'}{2}+a-{1\over2},\frac{\D'}{2}+b+{1\over2},\D'-h+1,1-\bz\bigg] \right.\nonumber\\
&&\hspace{5cm}\left.+ {}_2F_1\bigg[\frac{\D'}{2}+a+{1\over2},\frac{\D'}{2}+b-{1\over2},\D'-h+1,1-\bz\bigg]
\right.\nonumber\\
&&\hspace{4cm}\quad\quad\left.-{}_2F_1\bigg[\frac{\D'}{2}+a-{1\over2},\frac{\D'}{2}+b-{1\over2},\D'-h+1,1-\bz\bigg] \right) \, . 
\eeq
Each to those terms will contribute an hypergeometric function \eqref{anom1} with $\D'$ shifted accordingly.  Recalling the indentity,
\be
 {}_2F_1[a,b,c;z]=(1-z)^{-a}{}_2F_1\bigg[a,c-b,c; {z\over 1-z}\bigg]\,,
\ee
we can rewrite,
\beq
&& \, g_{\Delta,\,(1)} (a,b;(1-\bz),v)= \nonumber\\
&&  {1\over2} \,  \left( 
(1-\bz)^{-a} {}_2F_1\bigg[\frac{\D'}{2}+a-{1\over2},\frac{\D'}{2}-b+{1\over2}-h,\D'-h+1,{\bz\over 1-\bz}\bigg] \right.\nonumber\\
&&\hspace{5cm}\left.+ (1-\bz)^{-a-1}  {}_2F_1\bigg[\frac{\D'}{2}+a+{1\over2},\frac{\D'}{2}-b+{3\over2}-h,\D'-h+1,{\bz\over 1-\bz}\bigg]
\right.\nonumber\\
&&\hspace{4cm}\quad\quad\left.-(1-\bz)^ {-a}   {}_2F_1\bigg[\frac{\D'}{2}+a-{1\over2},\frac{\D'}{2}-b+{3\over2}-h,\D'-h+1,{\bz\over 1-\bz}\bigg] \right) \, . \nonumber\\
\eeq
which by putting it back into \eqref{smallzgen} gives a combination of three ${}_4F_3$ of a similar form as in the scalar block and the spin block in four dimensions. However we will not display the explicitly form because is large and we are not going to use it here. We just want to emphasize that it is possible to write down a closed form for the spining blocks in general dimension, even though as a large combination of Gauss hypergeometric functions.

\bibliographystyle{JHEP}
\bibliography{LSSCH}

\providecommand{\href}[2]{#2}\begingroup\raggedright\begin{thebibliography}{10}

\bibitem{Caron-Huot:2017vep}
S.~Caron-Huot, \emph{{Analyticity in Spin in Conformal Theories}},
  \href{http://dx.doi.org/10.1007/JHEP09(2017)078}{\emph{JHEP} {\bf 09} (2017)
  078}, [\href{https://arxiv.org/abs/1703.00278}{{\tt 1703.00278}}].

\bibitem{Rattazzi:2008pe}
R.~Rattazzi, V.~S. Rychkov, E.~Tonni and A.~Vichi, \emph{{Bounding scalar
  operator dimensions in 4D CFT}},
  \href{http://dx.doi.org/10.1088/1126-6708/2008/12/031}{\emph{JHEP} {\bf 12}
  (2008) 031}, [\href{https://arxiv.org/abs/0807.0004}{{\tt 0807.0004}}].

\bibitem{ElShowk:2012ht}
S.~El-Showk, M.~F. Paulos, D.~Poland, S.~Rychkov, D.~Simmons-Duffin and
  A.~Vichi, \emph{{Solving the 3D Ising Model with the Conformal Bootstrap}},
  \href{http://dx.doi.org/10.1103/PhysRevD.86.025022}{\emph{Phys. Rev.} {\bf
  D86} (2012) 025022}, [\href{https://arxiv.org/abs/1203.6064}{{\tt
  1203.6064}}].

\bibitem{Kos:2014bka}
F.~Kos, D.~Poland and D.~Simmons-Duffin, \emph{{Bootstrapping Mixed Correlators
  in the 3D Ising Model}},
  \href{http://dx.doi.org/10.1007/JHEP11(2014)109}{\emph{JHEP} {\bf 11} (2014)
  109}, [\href{https://arxiv.org/abs/1406.4858}{{\tt 1406.4858}}].

\bibitem{El-Showk:2014dwa}
S.~El-Showk, M.~F. Paulos, D.~Poland, S.~Rychkov, D.~Simmons-Duffin and
  A.~Vichi, \emph{{Solving the 3d Ising Model with the Conformal Bootstrap II.
  c-Minimization and Precise Critical Exponents}},
  \href{http://dx.doi.org/10.1007/s10955-014-1042-7}{\emph{J. Stat. Phys.} {\bf
  157} (2014) 869}, [\href{https://arxiv.org/abs/1403.4545}{{\tt 1403.4545}}].

\bibitem{Simmons-Duffin:2016wlq}
D.~Simmons-Duffin, \emph{{The Lightcone Bootstrap and the Spectrum of the 3d
  Ising CFT}}, \href{http://dx.doi.org/10.1007/JHEP03(2017)086}{\emph{JHEP}
  {\bf 03} (2017) 086}, [\href{https://arxiv.org/abs/1612.08471}{{\tt
  1612.08471}}].

\bibitem{Rychkov:2016iqz}
S.~Rychkov, \emph{{EPFL Lectures on Conformal Field Theory in D>= 3
  Dimensions}}.
\newblock SpringerBriefs in Physics. 2016,
  \href{http://dx.doi.org/10.1007/978-3-319-43626-5}{10.1007/978-3-319-43626-5}.

\bibitem{Simmons-Duffin:2016gjk}
D.~Simmons-Duffin, \emph{{The Conformal Bootstrap}},  in \emph{{Proceedings,
  Theoretical Advanced Study Institute in Elementary Particle Physics: New
  Frontiers in Fields and Strings (TASI 2015): Boulder, CO, USA, June 1-26,
  2015}}, pp.~1--74, 2017.
\newblock \href{https://arxiv.org/abs/1602.07982}{{\tt 1602.07982}}.
\newblock \href{http://dx.doi.org/10.1142/9789813149441_0001}{DOI}.

\bibitem{Poland:2018epd}
D.~Poland, S.~Rychkov and A.~Vichi, \emph{{The Conformal Bootstrap: Theory,
  Numerical Techniques, and Applications}},
  \href{https://arxiv.org/abs/1805.04405}{{\tt 1805.04405}}.

\bibitem{Komargodski:2012ek}
Z.~Komargodski and A.~Zhiboedov, \emph{{Convexity and Liberation at Large
  Spin}}, \href{http://dx.doi.org/10.1007/JHEP11(2013)140}{\emph{JHEP} {\bf 11}
  (2013) 140}, [\href{https://arxiv.org/abs/1212.4103}{{\tt 1212.4103}}].

\bibitem{Fitzpatrick:2012yx}
A.~L. Fitzpatrick, J.~Kaplan, D.~Poland and D.~Simmons-Duffin, \emph{{The
  Analytic Bootstrap and AdS Superhorizon Locality}},
  \href{http://dx.doi.org/10.1007/JHEP12(2013)004}{\emph{JHEP} {\bf 12} (2013)
  004}, [\href{https://arxiv.org/abs/1212.3616}{{\tt 1212.3616}}].

\bibitem{Alday:2015eya}
L.~F. Alday, A.~Bissi and T.~Lukowski, \emph{{Large spin systematics in CFT}},
  \href{http://dx.doi.org/10.1007/JHEP11(2015)101}{\emph{JHEP} {\bf 11} (2015)
  101}, [\href{https://arxiv.org/abs/1502.07707}{{\tt 1502.07707}}].

\bibitem{Alday:2015ota}
L.~F. Alday and A.~Zhiboedov, \emph{{Conformal Bootstrap With Slightly Broken
  Higher Spin Symmetry}},
  \href{http://dx.doi.org/10.1007/JHEP06(2016)091}{\emph{JHEP} {\bf 06} (2016)
  091}, [\href{https://arxiv.org/abs/1506.04659}{{\tt 1506.04659}}].

\bibitem{Alday:2015ewa}
L.~F. Alday and A.~Zhiboedov, \emph{{An Algebraic Approach to the Analytic
  Bootstrap}}, \href{http://dx.doi.org/10.1007/JHEP04(2017)157}{\emph{JHEP}
  {\bf 04} (2017) 157}, [\href{https://arxiv.org/abs/1510.08091}{{\tt
  1510.08091}}].

\bibitem{Kaviraj:2015cxa}
A.~Kaviraj, K.~Sen and A.~Sinha, \emph{{Analytic bootstrap at large spin}},
  \href{http://dx.doi.org/10.1007/JHEP11(2015)083}{\emph{JHEP} {\bf 11} (2015)
  083}, [\href{https://arxiv.org/abs/1502.01437}{{\tt 1502.01437}}].

\bibitem{Kaviraj:2015xsa}
A.~Kaviraj, K.~Sen and A.~Sinha, \emph{{Universal anomalous dimensions at large
  spin and large twist}},
  \href{http://dx.doi.org/10.1007/JHEP07(2015)026}{\emph{JHEP} {\bf 07} (2015)
  026}, [\href{https://arxiv.org/abs/1504.00772}{{\tt 1504.00772}}].

\bibitem{Simmons-Duffin:2017nub}
D.~Simmons-Duffin, D.~Stanford and E.~Witten, \emph{{A spacetime derivation of
  the Lorentzian OPE inversion formula}},
  \href{https://arxiv.org/abs/1711.03816}{{\tt 1711.03816}}.

\bibitem{Cardona:2018nnk}
C.~Cardona, \emph{{OPE inversion in Mellin space}},
  \href{https://arxiv.org/abs/1803.05086}{{\tt 1803.05086}}.

\bibitem{Gopakumar:2016wkt}
R.~Gopakumar, A.~Kaviraj, K.~Sen and A.~Sinha, \emph{{Conformal Bootstrap in
  Mellin Space}},
  \href{http://dx.doi.org/10.1103/PhysRevLett.118.081601}{\emph{Phys. Rev.
  Lett.} {\bf 118} (2017) 081601},
  [\href{https://arxiv.org/abs/1609.00572}{{\tt 1609.00572}}].

\bibitem{Dey:2017fab}
P.~Dey, K.~Ghosh and A.~Sinha, \emph{{Simplifying large spin bootstrap in
  Mellin space}}, \href{http://dx.doi.org/10.1007/JHEP01(2018)152}{\emph{JHEP}
  {\bf 01} (2018) 152}, [\href{https://arxiv.org/abs/1709.06110}{{\tt
  1709.06110}}].

\bibitem{Dey:2016mcs}
P.~Dey, A.~Kaviraj and A.~Sinha, \emph{{Mellin space bootstrap for global
  symmetry}}, \href{http://dx.doi.org/10.1007/JHEP07(2017)019}{\emph{JHEP} {\bf
  07} (2017) 019}, [\href{https://arxiv.org/abs/1612.05032}{{\tt 1612.05032}}].

\bibitem{Gopakumar:2016cpb}
R.~Gopakumar, A.~Kaviraj, K.~Sen and A.~Sinha, \emph{{A Mellin space approach
  to the conformal bootstrap}},
  \href{http://dx.doi.org/10.1007/JHEP05(2017)027}{\emph{JHEP} {\bf 05} (2017)
  027}, [\href{https://arxiv.org/abs/1611.08407}{{\tt 1611.08407}}].

\bibitem{Golden:2017fip}
J.~Golden and D.~R. Mayerson, \emph{{Mellin Bootstrap for Scalars in Generic
  Dimension}},  \href{https://arxiv.org/abs/1711.03980}{{\tt 1711.03980}}.

\bibitem{Dolan:2000uw}
F.~A. Dolan and H.~Osborn, \emph{{Implications of N=1 superconformal symmetry
  for chiral fields}},
  \href{http://dx.doi.org/10.1016/S0550-3213(00)00553-8}{\emph{Nucl. Phys.}
  {\bf B593} (2001) 599--633},
  [\href{https://arxiv.org/abs/hep-th/0006098}{{\tt hep-th/0006098}}].

\bibitem{Dolan:2000ut}
F.~A. Dolan and H.~Osborn, \emph{{Conformal four point functions and the
  operator product expansion}},
  \href{http://dx.doi.org/10.1016/S0550-3213(01)00013-X}{\emph{Nucl. Phys.}
  {\bf B599} (2001) 459--496},
  [\href{https://arxiv.org/abs/hep-th/0011040}{{\tt hep-th/0011040}}].

\bibitem{Dolan:2011dv}
F.~A. Dolan and H.~Osborn, \emph{{Conformal Partial Waves: Further Mathematical
  Results}},  \href{https://arxiv.org/abs/1108.6194}{{\tt 1108.6194}}.

\bibitem{Mack:2009mi}
G.~Mack, \emph{{D-independent representation of Conformal Field Theories in D
  dimensions via transformation to auxiliary Dual Resonance Models. Scalar
  amplitudes}},  \href{https://arxiv.org/abs/0907.2407}{{\tt 0907.2407}}.

\bibitem{Costa:2012cb}
M.~S. Costa, V.~Goncalves and J.~Penedones, \emph{{Conformal Regge theory}},
  \href{http://dx.doi.org/10.1007/JHEP12(2012)091}{\emph{JHEP} {\bf 12} (2012)
  091}, [\href{https://arxiv.org/abs/1209.4355}{{\tt 1209.4355}}].

\bibitem{Liu:2018jhs}
J.~Liu, E.~Perlmutter, V.~Rosenhaus and D.~Simmons-Duffin,
  \emph{{$d$-dimensional SYK, AdS Loops, and $6j$ Symbols}},
  \href{https://arxiv.org/abs/1808.00612}{{\tt 1808.00612}}.

\bibitem{Alday:2017zzv}
L.~F. Alday, J.~Henriksson and M.~van Loon, \emph{{Taming the
  $\epsilon$-expansion with Large Spin Perturbation Theory}},
  \href{https://arxiv.org/abs/1712.02314}{{\tt 1712.02314}}.

\bibitem{Cardona:2017tsw}
C.~Cardona, \emph{{Mellin-(Schwinger) representation of One-loop Witten
  diagrams in AdS}},  \href{https://arxiv.org/abs/1708.06339}{{\tt
  1708.06339}}.

\bibitem{Yuan:2018qva}
E.~Y. Yuan, \emph{{Simplicity in AdS Perturbative Dynamics}},
  \href{https://arxiv.org/abs/1801.07283}{{\tt 1801.07283}}.

\bibitem{Yuan:2017vgp}
E.~Y. Yuan, \emph{{Loops in the Bulk}},
  \href{https://arxiv.org/abs/1710.01361}{{\tt 1710.01361}}.

\bibitem{Giombi:2017hpr}
S.~Giombi, C.~Sleight and M.~Taronna, \emph{{Spinning AdS Loop Diagrams: Two
  Point Functions}},  \href{https://arxiv.org/abs/1708.08404}{{\tt
  1708.08404}}.

\bibitem{Bertan:2018khc}
I.~Bertan and I.~Sachs, \emph{{Loops in Anti-de Sitter Space}},
  \href{https://arxiv.org/abs/1804.01880}{{\tt 1804.01880}}.

\bibitem{Costa:2014kfa}
M.~S. Costa, V.~Goncalves and J.~Penedones, \emph{{Spinning AdS Propagators}},
  \href{http://dx.doi.org/10.1007/JHEP09(2014)064}{\emph{JHEP} {\bf 09} (2014)
  064}, [\href{https://arxiv.org/abs/1404.5625}{{\tt 1404.5625}}].

\bibitem{Sleight:2018epi}
C.~Sleight and M.~Taronna, \emph{{Spinning Mellin Bootstrap: Conformal Partial
  Waves, Crossing Kernels and Applications}},
  \href{https://arxiv.org/abs/1804.09334}{{\tt 1804.09334}}.

\bibitem{Sen:2017gfr}
K.~Sen and Y.~Tachikawa, \emph{{First-order conformal perturbation 
theory by marginal operators}},
\href{https://arxiv.org/abs/1711.05947}{{\tt 1711.05947}}. 

\end{thebibliography}\endgroup
\end{document}